\documentclass[aps,prb,twocolumn,showpacs,longbibliography,superscriptaddress,final,floatfix]{revtex4-1}
\usepackage{amsmath}
\usepackage{amsfonts}
\usepackage{amssymb}

\usepackage{color}

\usepackage{graphicx}
\usepackage{hyperref}
\usepackage{subfigure}
\usepackage{bbm}

\def\Xint#1{\mathchoice
	{\XXint\displaystyle\textstyle{#1}}%
	{\XXint\textstyle\scriptstyle{#1}}%
	{\XXint\scriptstyle\scriptscriptstyle{#1}}%
	{\XXint\scriptscriptstyle\scriptscriptstyle{#1}}%
	\!\int}
\def\XXint#1#2#3{{\setbox0=\hbox{$#1{#2#3}{\int}$}
		\vcenter{\hbox{$#2#3$}}\kern-.5\wd0}}

\def\dashint{\Xint-}

\newcommand{\new}[1]{{\color{black} #1}}
\newcommand{\neww}[1]{{\color{black} #1}}

\makeatletter

\begin{document}
	
	
	\title{Transfer matrix approach to quantum systems  subject to certain Lindblad evolution}
	
	
	\author{Junaid Majeed Bhat and Marko \v Znidari\v c\\\textit{Department of Physics, Faculty of Mathematics and Physics, \\University of Ljubljana, 1000 Ljubljana, Slovenia}}
	

	
	
	\date{\today}
	
	\begin{abstract}
          Solving for \neww{the} time evolution of a many particle system whose dynamics is governed by Lindblad equation is hard. We extend the use of the transfer matrix approach to a class of Lindblad equations that admit a closed hierarchy of two point correlators. An example that we treat is the XX spin chain, i.e., free fermions, subject to the local on-site dephasing, but can be extended to other Hermitian dissipators, e.g., non-local dephasing. We find a simple expression of the Green's function in the Laplace domain. The method can be used to get analytical results in the thermodynamic limit, for instance, to get the evolution of the magnetization density and to explicitly see the crossover between  ballistic and diffusive behavior, or to show that the correlations between operators at distance $l$ decay with time as $1/t^{\lceil l/2 \rceil+1/2}$. It also provides a fast numerical method to determine the evolution of the density with a complexity scaling with the system size more favorably than in previous methods, easily allowing one to study systems with $\sim 10^6$ spins.
	\end{abstract}
	
	\maketitle
	\section{Introduction}
	The transfer matrix formalism has been applied to solve wide variety of problems in physics. A few  notable examples include  propagation of electromagnetic waves, acoustic waves~\cite{yeh2012applied,pendry1992calculation,pendry1992universality},  waves of quantum particles such as electrons across any scatterers~\cite{mello2004quantum,beenakker1997random,kirkman1984statistics,ando1989numerical,pendry1982evolution,pendry1984transfer,Simple1981Lee1,Simple1981Lee2}, and computing partition functions~\cite{izergin1992determinant,schultz1964two}. This method applies when the solution of the problem can be built iteratively using  products of matrices. Therefore, it naturally becomes useful while dealing with problems requiring solutions of linear equations having finitely many non-zero off-diagonals, e.g. tri-diagonal or block tri-diagonal matrices~\cite{akaike1973block,molinari1997transfer,reuter2012efficient} which describe the Hamiltonian of non-interacting single particle systems.  Some simple examples of such systems  include the short-ranged  quantum tight-binding models or classical harmonic systems. For such systems the transfer matrix formalism allows one to obtain the elements of the single particle Green's function for the system~\cite{Simple1981Lee2,reuter2011probing}, which  determines properties such as the two-point correlation functions, current, conductance etc~\cite{dhar2006nonequilibrium,dhar2006heat}. 
	
	Another class of models where such problems occur are  quantum mechanical models in presence of dissipation or dephasing due to  certain  type of Hermitian Lindblad operators~\cite{vznidarivc2010exact,zunk_2014,barthel2022solving,wang2024,vznidarivc2011solvable,vznidarivc2024superdiffusive,eisler2011crossover,temme2012stochastic}. In these cases, the dynamics of two point correlation matrix is given by a linear differential equation where the matrix governing the dynamics has finitely many non-zero off-diagonals. One should therefore be able to apply the transfer matrix method in a manner  similar to tight-binding lattices to obtain solutions for such equations, and this is the approach taken in this paper. The transfer matrix   approach avoids the  calculation of the entire \new{spectral decomposition} of the matrix governing the dynamics of the correlators \new{and instead expresses the solution as a power of a simple transfer matrix of some small dimension}. The latter allows one to take the thermodynamic limit with ease.  In this paper, we demonstrate the transfer matrix approach in the simple case of XX spin chain  with $L$ spins in \neww{the} presence of local dephasing and show that this approach leads to a simple solution for the two-point correlation matrix. We express the solution for the correlators in terms of $L^{th}$ power of a $2\times 2$ transfer matrix, and therefore the thermodynamic limit just requires determining its largest eigenvalue and eigenvector. \new{Once the correlation matrix is known, its elements determine physical observables of general interest. For example, the diagonal elements of the correlation matrix directly give the magnetization density and the imaginary parts of the first off-diagonal elements give the magnetization current between different bonds in the chain. Using the transfer matrix approach, we express the Green's function   for the magnetization density   as a rather simple function in the Laplace domain whose analytic structure determines \neww{the} type of the behavior we observe.  At long times diffusion arises due to contribution from real poles of the Green's function while at short times the ballistic behavior comes from the contribution of its  branch cut. 
	} 
	
	XX spin chain with dephasing has been extensively studied in the literature, e.g. Refs.~\onlinecite{vznidarivc2010exact,medvedyeva2016exact,Alex2024,silva2023nontrivial,caoSciPost19,turkeshiPRB21,vznidarivc2015relaxation,carolloPRE17,bernardPRL19,hagaPRR23}, and therefore serves as an ideal simple case that one can consider, and where one can study e.g. a crossover from ballistic magnetization transport at short times to diffusive behavior at long times. \neww{One can think of it as the simplest solvable model displaying such a transition.} In the continuum limit, a qualitative understanding of the crossover can be described by considering two conserved quantities, say energy and particle density, that are coupled through a scattering. In our spin language we will have the magnetization density  $m(x)$ and the magnetization current $j(x)$, with the simplest possible two coupled continuity equations being,
	\begin{align}
		\partial_t{m}(x,t)&=-\partial_x j(x,t) \label{eq:te1}\\
		\partial_t{j}(x,t)&= -v^2 \partial_x m(x,t)-4\gamma j(x,t)\label{eq:te2},
	\end{align}
where $v$ is the velocity and $\gamma$ the dephasing strength due to scattering (decay rate of the current). This set of equations is equivalent to the 2nd order equation, 
	\begin{equation}
	\partial_{xx}m(x,t)=\frac{1}{v^2}\partial_{tt} m(x,t)+\frac{4\gamma}{v^2} \partial_t m(x,t),\label{eq:te3}
	\end{equation}
        which is nothing but the so called telegrapher's equation, commonly used in signal analysis for transmission and propagation of electrical signals~\cite{metaxas1983industrial}. For small times the diffusive time derivative can be neglected and the equation will behave as a wave equation with velocity $v^2$, while for long times the second time derivative is negligible and it goes into a diffusion equation with diffusion constant $D=v^2/(4\gamma)$ (see e.g. Ref.~\onlinecite{kac74} for \neww{a simple approach} to solve telegrapher's equation). \neww{Another way to look at it is that it corrects violation of causality in diffusion equation by introducing relaxation time~\cite{Kadanoff} (Green's function for diffusion equation is nonzero everywhere in space already for arbitrarily small times).} As we shall see, the equations for the XX chain with dephasing are quite similar, and one can view it as a lattice version of the telegrapher's equation with a tight-binding dispersion relation, i.e. \neww{there are particles with different velocities.} What we will get out of our analysis is a behavior that continuously transitions with time from ballistic to diffusive transport, or, in mathematical terms in the continuum limit from a hyperbolic to a parabolic partial differential equation.

	This paper is structured as follows:  In sec.~\ref{sec:model}, we introduce the model and the matrix equation for the two point correlators that we want to solve. In the next section, sec.~\ref{sec:tma}, we adopt the transfer matrix approach to obtain the solution for the correlation matrix at time $t$ for any finite size $L$. In the following section, sec.~\ref{sec:tl} we look at the solution in the thermodynamic limit and discuss the asymptotic  behavior of \neww{the} diagonal and the off-diagonal elements of the correlation matrix. We also present numerical results for the transferred magnetization and its logarithmic derivative \neww{(finite-time transport dynamical exponent)} and discuss the crossover from the ballistic to diffusive behavior. We conclude in section ~\ref{sec:concl}.

 \section{XX-Model with Dephasing}
 \label{sec:model}
In this section, we introduce the model and setup \neww{known} equations for the two point correlators that we wish to solve. We consider an $XX$ spin chain with periodic boundary conditions. The Hamiltonian is given by,
\begin{equation}
  H=-J\sum_{j=0}^{L-1} \sigma^j_1\sigma^{j+1}_1+\sigma^j_2\sigma^{j+1}_2,
\end{equation}
where $J$  is the hopping strength between the nearest neighbor sites and $\sigma_1^j,~\sigma_2^j,~\sigma_3^j$ \new{are Pauli spin operators for the $j^{th}$ spin defined as follows,}
\begin{equation}
	\sigma_1^j=\begin{pmatrix}
		0 && 1\\
		1 &&0
	\end{pmatrix},~\sigma_2^j=\begin{pmatrix}
	0 && -i\\
	i &&0
\end{pmatrix},~\sigma_3^j=\begin{pmatrix}
1 && 0\\
0 &&-1
        \end{pmatrix}.
\end{equation}
The chain is subjected to dephasing due to Lindblad operators  given by $L_j=\sqrt{\gamma/2}\sigma^j_3$, $\gamma$ is the dephasing strength, which act on each spin. The dynamics of the density matrix of the system is therefore given by the Lindblad master equation,
\begin{equation}
	\dot{\rho}=i [\rho,H]+\sum_{j=0}^{L-1}\left([L_j, \rho L_j^\dagger]+[L_j\rho,  L_j^\dagger]\right).
\end{equation}

Solving the master equation directly  requires solving  $4^L-1$ coupled differential equations, and is therefore hopeless for large system sizes. However, certain~\cite{eisler2011crossover,zunk_2014,barthel2022solving} Hermitian Lindblad operators with quadratic Hamiltonians allow reducing the exponential complexity in system size to a polynomial complexity. This is achieved by decoupling blocks of observables based on the number of Fermionic operators they contain resulting in a hierarchy of equations. In this hierarchy, lower point correlators serve as source terms for the equations of higher point correlators. For example, the block of two point correlators decouples from the higher point correlators and the two point correlators satisfy a closed set of linear equations~\cite{vznidarivc2010exact,vznidarivc2011solvable,zunk_2014,barthel2022solving,eisler2011crossover}. These then serve as source terms for third order correlators and so on.  

 We are interested only in the dynamics of the two point correlators in this paper. Let us define a \new{Hermitian two point correlation matrix as,
 \begin{equation}
 	C_{jk}(t)=\begin{cases}
 		 \langle A_j^{(k-j+1)}(t)\rangle+i \langle B_{j}^{(k-j+1)}(t)\rangle,~ k>j\\
 		 \langle A_j^{(1)}(t)\rangle,~k=j\\
 		 C_{jk}^*(t),~k<j
 	\end{cases}
 \end{equation}
 where, $\langle O(t) \rangle=Tr[\rho(t) O]$ denotes the expectation value of an operator $O$ at time $t$, and the operators $A_j^r(t)$ and $B_j^r(t)$ are  given by,
 \begin{align}
 	A_j^{(r)}(t)&=\sigma_1^j Z_{j+1}^{(r-2)}\sigma_1^{j+r-1}+\sigma_2^j Z_{j+1}^{(r-2)}\sigma_2^{j+r-1},\\
 	B_j^{(r)}(t)&=\sigma_1^j Z_{j+1}^{(r-2)}\sigma_2^{j+r-1}-\sigma_2^j Z_{j+1}^{(r-2)}\sigma_1^{j+r-1}.
 \end{align}
}
 $Z_j^{(r)}=\sigma_3^j...\sigma_3^{j+r-1}$ defines a string of $\sigma_3$ operators \neww{($Z_j^{(0)}=\mathbbm{1}$)}, and $A_j^{(1)}(t)=-\sigma_3^j(t)$. 
It can be shown~\cite{vznidarivc2013transport} that the dynamics of the two point correlation matrix is given by the equation,
\begin{equation}
  \frac{d}{dt} C(t)= -2 i [\mathbf{T}C(t)-C(t)\mathbf{T}^T]-2[\mathbf{\Gamma}\tilde{C}(t)+\tilde{C}(t)\mathbf{\Gamma}],\label{eq:correq}
\end{equation}
where $\mathbf{T}_{jk}=J(\delta_{j,k-1}+\delta_{j,k+1})$, and $\mathbf{T}_{1,L}=\mathbf{T}_{L,1}=J$ for our model with only nearest neighbor couplings.  $\mathbf{\Gamma}_{jk}=\gamma \delta_{jk}$, and $\tilde{C}(t)=C-\text{diag}[C(t)]$.  \new{The  elements of the correlation matrix give different physical observables that one may be interested to study. For example, the diagonal elements of $C$, $C_{jj}=-\langle\sigma_3^j\rangle$ give the magnetization density of the spin chain and imaginary parts of the elements of the first off-diagonal i.e. $C_{j,j+1}$ gives the current on the bond between $j$ and $j+1$.}

Eq.~(\ref{eq:correq}) will be the central focus of our analysis and in the next section we  show that this equation yields a simple solution via the transfer matrix approach. While we consider a simple case of homogeneous and nearest neighbor couplings, one can carry out the same approach for Hamiltonians with  different couplings for odd and even bonds \neww{and} for other Hermitian Lindblad operators such as the nonlocal dephasing of Ref.~\onlinecite{wang2024} and Ref.~\onlinecite{vznidarivc2024superdiffusive} where the two point correlators still satisfy a closed set of linear equations of the type similar to Eq.~(\ref{eq:correq}).

 \section{Transfer Matrix Approach}
 \label{sec:tma}
In this section, we develop a solution for the evolution of  the correlation matrix using the transfer matrix approach. 
We begin by considering the equation for the correlators in the component form,
 \begin{align}
 	&\frac{d }{dt}C_{x,y}= -2 i J (C_{x+1,y}+C_{x-1,y})\notag+2i J(C_{x,y+1}+C_{x,y-1})\\&~~~~~~~~~~~~~-4 \gamma C_{x,y}+\delta_{x,y}4 \gamma C_{x,y}.\label{eq:Cxydot}
 \end{align}
 We Fourier transform from $x$ to \new{momentum} $q_n=2\pi n/L$, $n=1, 2,...,L$, to exploit the periodic boundaries, as follows
 \begin{equation}
 	C_{x,y}(t)= \frac{1}{L}\sum_{n=1}^{L} e^{iq_n x} i^{l} e^{-i q_n l/2 } g_l(t,q_n),\label{eq:ctog}
 \end{equation}
where $l=x-y$ \new{such that $l=-(L-1),-(L-2),...,-1,0,1,...,L-1$.  Using Eq.~(\ref{eq:ctog}) in  Eq.~(\ref{eq:Cxydot})}   we obtain,
 \begin{align}
 	\frac{d}{dt} g_l(t,q)
 	&=\sum_{l'=-(L-1)}^{L-1}\mathcal{A}_{ll'} g_{l'}(t,q),\label{eq:dotgl}
 \end{align}
 where $\mathcal{A}$ is a   matrix with  components given by 
 \begin{equation}
 	\mathcal{A}_{ll'}=i\frac{\omega(q)}{2}(\delta_{l,l'+1}+\delta_{l,l'-1})+4\gamma (\delta_{l,0}-1)\delta_{l,l'},
 \end{equation}
with 
\begin{equation}
\omega(q)=8J\sin(q/2).
\end{equation} 

 The set of the equations in Eq.~(\ref{eq:dotgl}) are the same, up to a re-parametrization, as the ones obtained in Ref.~\onlinecite{eisler2011crossover} which considers a XX spin chain subject to incoherent hoppings. However, the solution was obtained by obtaining the  spectral decomposition of $\mathcal{A}$ which gives,
\begin{equation}
	g_l(t,q_n)=\sum_{\nu,l'} e^{ \lambda_\nu t} U^{-1}_{l,\nu} U_{\nu,l'}g_{l'}(0,q_n), \label{eq:sdeq}
\end{equation}
where $\lambda_\nu$ are the eigenvalues of $\mathcal{A}$ and $U$ consists of the corresponding eigenvectors of $\mathcal{A}$. Note that in this solution the dependence on the system size is implicit in the behavior of eigenvectors and the eigenvalues with the system size $L$. Therefore, taking $L\rightarrow\infty$ requires finding the limiting behavior of the eigenvalues and the components of the eigenvectors which in Ref.~\onlinecite{eisler2011crossover} reduces down to solving a rather complicated transcendental  equation for $L\rightarrow\infty$.   We will see that  the transfer matrix approach avoids the cumbersome  calculation of the entire spectrum and the eigenvectors of $\mathcal{A}$ and gives  a solution where the dependence on the system size is explicit due to which the limit $L\rightarrow\infty$ can be carried out with ease. It also provides  the solution for the full correlation matrix in the limit $L\rightarrow\infty$ whereas Ref.~\onlinecite{eisler2011crossover} considers only the diagonal elements of the correlation matrix in this limit.

 We  consider Eq.~(\ref{eq:dotgl}) only for $l>0$ as $C_{x,y}(t)$ is Hermitian. Note that this equation simply corresponds to   a nearest neighbor tight-binding lattice with the hopping strength of $\omega(q)/2$, and an onsite \new{imaginary} chemical potential  of $4\gamma$ present at all sites except at $l=0$.  \new{Therefore, the standard transfer matrix formalism~\cite{Simple1981Lee1,Simple1981Lee2} for tight-binding systems can be applied. However, this formalism is applied to tight-binding systems with boundaries\neww{. Therefore,} Eq.~(\ref{eq:dotgl}) for $l=0$,} which reads
 \begin{align}
 	\frac{d }{dt}g_0(t, q_n)&= i \frac{\omega(q_n)}{2}(g_1(t,q_n)+g_{-1}(t,q_n)),\label{eq:g01}
 \end{align}
needs to be modified such that it does not contain $g_{-1}(t,q_n)$.  For this purpose, we exploit another symmetry allowed by the equations for the correlators. We note that if the initial condition $C_{xy}(0)$ is such that the even diagonals of the correlation matrix are real and the odd diagonals are imaginary, then this is true for any time \new{(See appendix \ref{app:symm})}.   
\new{The simplest initial state which guarantees this structure is the initial state where the system is  in a product state in $\sigma^z$ basis, i.e. $\rho(0)$ is a pure state given by  $\rho(0)=|\psi_0\rangle\langle\psi_0|$ where 
\begin{equation}
	|\psi_0\rangle=|b_1\rangle\otimes|b_2\rangle...|b_{L-1}\rangle\otimes|b_L\rangle,\label{eq:initstate}
\end{equation}
$b_x=\pm$, and $\sigma_z^x|b_x\rangle=b_x|b_x\rangle$, i.e. $|\pm\rangle$ are the usual up and down states for the spin. Clearly, for this} initial state $C_{x,y}(0)$ is diagonal i.e. $C_{x,y}(0)=\delta_{x,y}c_x$, and for the rest of this paper we fix this to be our initial condition. 

 Using the symmetry allowed by our choice of the initial state, we have $C_{x+1,x}=C_{x,x+1}^*=-C_{x,x+1}$, which gives $g_1(t,q_n)=g_{-1}(t,q_n)$ and thus for $l=0$ we have,
 \begin{align}
 	\frac{d}{dt} g_0(t,q_n)=i\omega(q_n)g_{1}(t,q_n).\label{eq:dotg0}
 \end{align}
 Putting together Eq.~(\ref{eq:dotg0}) for $g_0$ and \new{ equations for $g_l$ with $l>0$ from Eq.~(\ref{eq:dotgl}), we have
 \begin{align}
 	\dot g_l(t,q_n)&= \sum_{l'=0}^{L-1}A_{ll'} g_{l'}(t,q_n),\label{eq:gtq}
 \end{align}
which gives
\begin{align}	
 	 g_l(t,q_n)&= \sum_{l'=0}^{L-1}G_{ll'}(t) g_{l'}(0,q_n)\label{eq:gtq1},
 \end{align}
 where  $A$ is a tridiagonal matrix as follows,
 \begin{equation}
 	A=\begin{bmatrix}
 		0 & i\omega(q_n) & 0 & \cdots & 0 \\
 		i\omega(q_n)/2 & -4\gamma & i\omega(q_n)/2 & \ddots & \vdots \\
 		0 & i\omega(q_n)/2 & -4\gamma & \ddots & 0 \\
 		\vdots & \ddots & \ddots & \ddots &i\omega(q_n)/2  \\
 		i\omega(q_n)/2 & \cdots & 0 & i\omega(q_n)/2 & -4\gamma
 	\end{bmatrix},
 \end{equation}
and $G(t)=e^{A t}$.}

 To apply the transfer matrix formalism, we  take a Laplace transform~\footnote{A Fourier transform instead of a Laplace transform can also be used. However, one needs to add a small negative real part to the eigenvalues of $A$ as the Green's function $(i\omega-A)^{-1}$ will diverge if $A$ has imaginary eigenvalues. To avoid this, we simply stick to the Laplace transform which does not have such problems.} of Eq.~(\ref{eq:gtq1})  to get,
 \begin{equation}
 	\tilde g_l(s,q_n)= \sum_{l'=0}^{L-1}\mathcal{G}_{ll'}(s)g_{l'}(0,q_n),\label{eq:tildegl}
 \end{equation}
 where $\tilde{g_l}(s,q_n)=\int_0^\infty dt e^{-st} g_l(t,q_n)$ is the Laplace transform of $g_l(t,q_n)$, \new{and the matrix  
 	\begin{equation}
 		\mathcal{G}(s)=\frac{1}{s-A}
 	\end{equation} 
 defines Green's function in the Laplace domain. We will see shortly that for product initial states of the type in Eq.~(\ref{eq:initstate}), only the first \neww{element} of the matrix $\mathcal{G}(s)$, $\mathcal{G}_{0,0}(s)$, determines the entire solution and can be expressed \neww{in terms of} a product of a simple $2\times2$ transfer matrices. 
} 	  
 
 Let us first consider the solution for the diagonal elements of $C(t)$ and consider the remaining elements at the end of this section. The solution for diagonal elements  can be written in terms of $\tilde g_0(s,q_n)$ as,   
 \begin{equation}
   C_{x,x}(t)=\frac{1}{L} \sum_{n=1}^{L} e^{iq_n x}  \mathcal{L}^{-1}[\tilde g_0(s,q_n)],\label{eq:ctotildeg}
\end{equation}
where $\mathcal{L}^{-1}$ \new{stands for the  inverse Laplace transform defined as,
\begin{equation}
	\mathcal{L}^{-1}[f(s)]=\int_{\eta-i\infty}^{\eta+i\infty} \frac{ds}{2\pi i} e^{st} f(s)\label{eq:lapinv}
\end{equation}
where $\eta$ is taken such that it is greater than the real parts of all the singularities of the function $f(s)$. } As the initial condition is diagonal $g_l(0,q_n)=\delta_{l,0} c(q_n)$,  $c(q_n)$ is the Fourier transform of $c_x$ i.e.
 \begin{equation}
   c(q_n)=\sum_{x=0}^{L-1} e^{-i q_n x} c_x,
 \end{equation} 
which gives
\begin{equation}
\label{eq:g0tilde}	\tilde g_{0}(s,q_n)=\mathcal{G}_{0,0}(s)c(q_n).
\end{equation}
 Using Eq.~(\ref{eq:g0tilde}) in Eq.~(\ref{eq:ctotildeg}) we see that the diagonal elements are determined by the first element $\mathcal{G}_{0,0}(s)$ of  the matrix $\mathcal{G}(s)$.
 
   \neww{To express $\mathcal{G}_{0,0}(s)$ in terms of a product of $2\times2$  matrices, we consider the  first column of equations from the identity} \begin{equation}
   		(s-A)\mathcal{G}(s)=I,\label{eq:Giden}
	\end{equation}
   	namely Eq.~(\ref{aeq:ideq1}-\ref{aeq:ideq3}). We show in  appendix~\ref{app:tfm} that these equations can be used   to obtain the following relation,
 \begin{align}
 	\begin{pmatrix}
 		1\\\mathcal{G}_{0,0}
 	\end{pmatrix}&=T_0 T^{L-1}\begin{pmatrix}
 		\mathcal{G}_{L-1,0}\\\mathcal{G}_{0,0}
 	\end{pmatrix}.\label{eq:g0gN}
 \end{align}
$T_0$ and $T$ are the transfer matrices which are given by,
\begin{align}
	T_0=\begin{pmatrix}s & -i\omega(q_n)\\
		1 & 0
	\end{pmatrix}~\text{and}~T=	\begin{pmatrix}
	-2i u & -1 \\
	1 & 0
\end{pmatrix}
\end{align}
where $u=(s+4\gamma)/\omega(q_n)$. \new{ $T_0$ is the boundary transfer matrix which arises due to the fact that equation for $l=0$ is different from equations for $l>0$ in Eq.~(\ref{eq:gtq}), and $T$ is the transfer matrix in the bulk.} 

  Eq.~(\ref{eq:g0gN}) gives two linear equations for $\mathcal{G}_{00}$ and $\mathcal{G}_{L-1,0}$ which can be easily solved to obtain that
  \begin{align}
 	\mathcal{G}_{0,0}=\bigg[\frac{\tau_L}{\sin(\alpha L)}+\frac{1}{\sin(\alpha)}\bigg[-\tau_{L-1}+\frac{\tau_L \sin(\alpha(L-1))}{\sin(\alpha L)}\bigg]\bigg]^{-1},\label{eq:g00L}
 \end{align}
 where $\tau_L=s\sin(\alpha L)-i\omega(q_n)\sin(\alpha(L-1))$, and $\alpha=\arccos(-i u)$.
 
 \subsection{Other Models}
  Due to the simplicity of the model, we were able to get a simple expression for the required \neww{element of the matrix} $\mathcal{G}(s)$ for any finite $L$ and of-course \new{explicit solutions such as in Eq.~(\ref{eq:g00L})} \neww{are not always} possible. Nevertheless, the product structure of Eq.~(\ref{eq:g0gN}) is always  possible to obtain as  long as there are finitely many non-zero off-diagonal elements in $A$, and the solution is such that even and odd diagonals are purely real and complex, respectively. For example, for models with different hopping  on even and odd sites, the transfer matrix will still be $2\times2$ however the product structure will now contain two different matrices occurring alternately. The dimensions of the transfer matrix is determined by the non-locality of the couplings present and therefore for   models with couplings only up to a few neighbors, the dimensions of the transfer matrix will be of the same order.  For example,  the non-local dephasing model with three site Lindblad operators~\cite{wang2024,vznidarivc2024superdiffusive} (resulting in superdiffusion) defined as $L_j=l_j^\dagger l_j$ where,
  \begin{equation}
  	l_j=\sqrt{\frac{\gamma}{2}}(\sigma_{j-1}^-+Z^{(2)}_{j-1}\sigma_{j+1}^{-}),
  \end{equation}
results in a transfer matrix \new{in the bulk} of the form 
\begin{equation}
\begin{pmatrix}
 		a_1(q) && a_2(q) && a_3(q) && a_4(q)\\
 		1 &&0 && 0 &&0 \\
 		0 &&1 && 0 &&0 \\
 		0 &&0 && 1 &&0 \\
 	\end{pmatrix},
 \end{equation}
 where $a_1(q),~a_2(q),~a_3(q),~a_4(q)$ are defined as
 \begin{align}
	&a_1(q)=\frac{2 iJ}{\gamma}\frac{1-e^{i q}}{e^{2iq}+1},~a_2(q)=\frac{1}{\gamma}\frac{2\gamma-s}{1+e^{2i q}}\notag\\&~a_3(q)=\frac{2 iJ}{\gamma}\frac{1-e^{-i q}}{e^{2iq}+1},~a_4(q)=\frac{e^{-2i q}-1}{e^{2iq}+1}.
\end{align}
 Once the transfer matrix is known, behavior of the product can be studied numerically  for large and small $s$ limit  for thermodynamically large system sizes, which as we will see in the next section determines the asymptotic behavior of the correlators at short and long times, respectively. 
 
 \subsection{Off-diagonal elements and non-diagonal initial states}
 \neww{We now obtain} the solution for the off-diagonal elements of $C(t)$ in terms of $\mathcal{G}_{0,0}$. The \new{$l^{th}$} off-diagonal elements can be written as,
 \begin{equation}
 	C_{x+l,x}(t)=\frac{1}{L} \sum_{n=1}^{L} e^{iq_n x} i^l e^{-i q_n l/2} \mathcal{L}^{-1}[\tilde g_l(s,q_n)],\label{eq:cltotildeg}
 \end{equation}
 where $\tilde g_l(s,q_n)=\mathcal{G}_{l,0}c(q_n)$. The components $\mathcal{G}_{l,0}$ can be determined using the same iteration procedure which we used to obtain Eq.~(\ref{eq:g0gN}). One obtains the following
 \begin{align} 
 	\begin{pmatrix}
 		1\\\mathcal{G}_{0,0}
 	\end{pmatrix}&=T_0 T^{l}\begin{pmatrix}
 		\mathcal{G}_{l,0}\\\mathcal{G}_{l+1,0}
 	\end{pmatrix}\label{eq:g0gl}
 \end{align}
 Since $\mathcal{G}_{0,0}$ is already known, the above equations can be inverted to get $\mathcal{G}_{l,0}$ to be,
 \begin{equation}
 	\mathcal{G}_{l,0}=\frac{-1}{i \omega\sin \alpha}\left[\sin(\alpha l)+(i \omega \sin[\alpha(l-1)]-s\sin \alpha l)\mathcal{G}_{0,0}\right]\label{eq:gl0}.
 \end{equation}

\neww{Note the fact that only the elements of the first column of the matrix $\mathcal{G}$ determine the entire correlation matrix is a direct consequence of choosing the initial correlation matrix to be diagonal. For initial states with non-diagonal correlation matrix, it follows from Eq.~(\ref{eq:tildegl}) that one needs to determine the  elements of other columns of the matrix $\mathcal{G}$. For instance, in presence of nonzero currents, i.e. nonzero first off-diagonal of the initial correlation matrix,  elements of the second column of $\mathcal{G}$ are also required. These can be determined using the same transfer matrix procedure but now considering the second column of equations from the identity in Eq.~(\ref{eq:Giden}).}

 
\section{Thermodynamic limit and Asymptotic behavior}
\label{sec:tl}
\subsection{Thermodynamic Limit}
In the last section, we showed that  the solution for all the correlators  is basically determined by a single function $\mathcal{G}_{0,0}$ whose dependence on $L$ is explicitly given in Eq.~(\ref{eq:g00L}).  So, taking the thermodynamic limit is taking $L\rightarrow\infty$ for this function. In this limit, we have $\sin(\alpha L)\approx-\frac{e^{-i \alpha L}}{2i}$ since $\alpha=\arccos(-iu)$ has a positive imaginary part as $u>0$. \new{Substituting $\sin(\alpha L)\approx-\frac{e^{-i \alpha L}}{2i}$ in Eq.~(\ref{eq:g00L})  $\mathcal{G}_{0,0}$ simplifies to,
 \begin{align}
 \mathcal{F}(\tilde s,q)&=\lim\limits_{L\rightarrow\infty}\mathcal{G}_{0,0}\\&=\frac{1}{\tilde{s}-4\gamma-i\omega(q) e^{i\alpha}}=\frac{1}{\sqrt{\tilde s^2+\omega(q)^2}-4\gamma}\label{eq:G00inf},
 \end{align}}
where $\tilde s=s+4\gamma$ and we used the fact that 
 \begin{align}
\alpha= \arccos(-iu)=\frac{\pi}{2}+i \log\big[u+\sqrt{u^2+1}\big]~\label{eq:arcos}.
 \end{align}
 \neww{The solution for diagonal elements  in the limit $L\rightarrow\infty$  now reads,
 \begin{align}
 C_{x,x}(t)=\int_0^{2\pi} \frac{dq}{2\pi} e^{iq x}e^{-4\gamma t} \mathcal{L}^{-1}[\mathcal{F}(s, q)] c(q),\label{eq:Cxxfeq}
 \end{align}
 where we have converted the sum over $q_n$ into an integral.  Note that the thermodynamic limit naturally came out of the solution in Eq.~(\ref{eq:g00L}), this is one advantage of using this method as opposed to taking the thermodynamic limit via the spectral decomposition of $\mathcal{A}$, i.e. Eq.~(\ref{eq:sdeq}).
 
 For initial states in Eq.~(\ref{eq:initstate}), ignoring the long range correlations $l\geq 2$  and denoting $C_{x,x}=-m(x)$ and $C_{x+1,x}=i j(x)/(4J)$, the resulting equation for $m(x)$ and $j(x)$ in the continuum limit are the same  as  Eq.~(\ref{eq:te1}) and Eq.~(\ref{eq:te2}), respectively, with $v^2=8 J^2$. These lead to the telegrapher's equation for the magnetization density, $m(x)$, and thus we expect a diffusion constant of $2 J^2/\gamma$ in the long time limit.  This must also follow from expanding $\mathcal{F}(\tilde s,q)$ from Eq.~(\ref{eq:G00inf}) in the long wavelength limit i.e. around $q=0$. After doing so, we get
\begin{equation}
	\mathcal{F}^{te}(s,q)=\frac{s+4\gamma}{s^2+4\gamma s+ 8(J q)^2}
\end{equation}
which is indeed the Green's function for the telegrapher's equation,  Eq.~(\ref{eq:te3}), with $v^2=8J^2$, as expected. While the telegrapher's equation gives the correct diffusive behavior in the long time limit, it does not  not give the correct relaxation to the diffusive behavior as it ignores the dispersion, $\omega(q)$, of the system.
}

 \subsection{Asymptotic Behaviors and  Analytic structure}
  Eq.~(\ref{eq:Cxxfeq})  provides a simple and useful expression for the density for following reasons. The long time and short time behavior follow very elegantly by simply looking at this equation in small $s$ and large $s$ limit, respectively. Let us  look at these two limits separately and for simplicity \neww{ we consider  $c(q)=1$ i.e. an initial state which gives the magnetization density to be $c_x=\delta_{x,0}$.}

\textit{Short time limit:} In this limit we consider $s\sim \omega(q) > 4\gamma $, so dropping $4\gamma$ in the denominator of Eq.~(\ref{eq:Cxxfeq}) we get,
\begin{align}
&C_{xx}(t)\sim\int_0^{2\pi} \frac{dq}{2\pi} e^{iq x}e^{-4\gamma t} \mathcal{L}^{-1}\Bigg[\frac{1}{\sqrt{s^2+\omega^2(q)}}\Bigg]\\&=e^{-4\gamma t}\int_0^{2\pi} \frac{dq}{2\pi} e^{iq x} J_0[\omega(q) t]= [J_x(4Jt)]^2 e^{-4\gamma t} ,\label{eq:Cxxlarges}
\end{align}
where the Bessel function, $J_0[\omega(q) t]$ arises from the inverse Laplace of $1/\sqrt{s^2+\omega^2(q)}$, and the square of the Bessel  $J_x(4Jt)$, known to be the Green's function in absence of  dephasing~\cite{economou2006green}, comes by the subsequent integration over $q$. We see that   the ballistic behavior starts disappearing exponentially as soon as the dephasing is turned on.

\textit{Long time limit:} In this case, we first rewrite Eq.~(\ref{eq:Cxxfeq}) by re-shifting the factor $e^{-4\gamma t}$ back into the Laplace inverse integral which gives,
\begin{align}
C_{xx}(t)=\int_0^{2\pi} \frac{dq}{2\pi} e^{iq x} \mathcal{L}^{-1}\Bigg[\frac{1}{\sqrt{(s+4\gamma)^2+\omega^2(q)}-4\gamma}\Bigg].\label{eq:Cxxfeq1}
\end{align}
For the long time limit, we  consider  $4\gamma>s\sim \omega(q)$. Also, since the behavior is dominated by small $q$, i.e. long wavelengths, $\omega(q)\sim 4 J q$. Using this and extending the limits in the $q$ integral from $-\infty$ to $\infty$ as the dominant contribution  comes near $q=0$, we have
\begin{align}
C_{xx}(t)\approx\int_{-\infty}^{\infty}\frac{dq}{2\pi}e^{iq x}  \mathcal{L}^{-1}\Bigg[ \frac{1}{s+(2 J^2 q^2/\gamma)}\Bigg]=\frac{e^{-x^2/(4Dt)}}{\sqrt{4\pi D t}},\label{eq:Cxxsmalls}
\end{align}
where $D=2J^2/\gamma$ is the diffusion constant which is in accordance with the expectations from the telegrapher's equation. Note that the small $s$ and $q$ behavior of $\mathcal{G}_{0,0}$ is given by $(s+(2 J^2 q^2/\gamma))^{-1}$, and its Laplace inverse gives  the diffusion kernel, i.e. $e^{-2 J^2q^2t/\gamma}$, as expected. A deviation from this behavior in $\mathcal{F}(s, q)$   would give rise to anomalous behavior.

\begin{figure}[h!]
	\centering
	\subfigure{
          \includegraphics[width=0.23\textwidth,page=1]{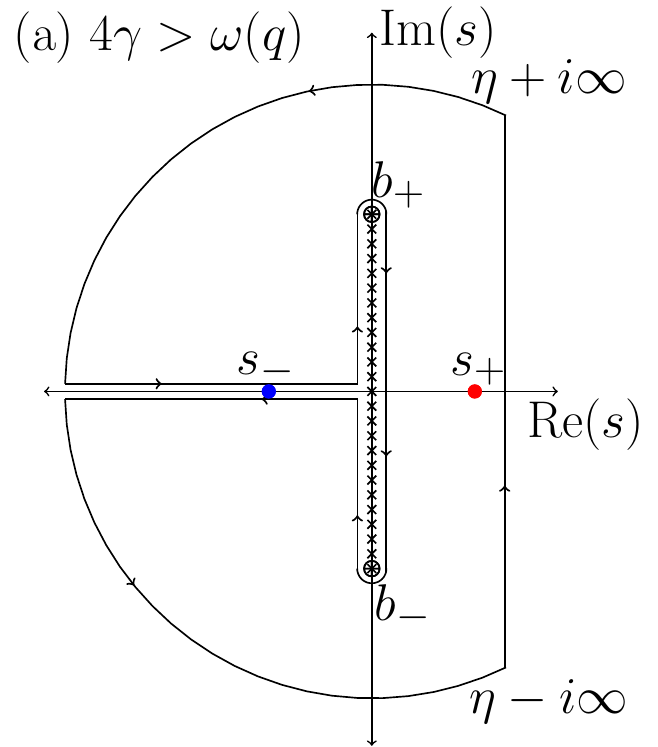}}
	\subfigure{
		\includegraphics[width=0.23\textwidth,page=2]{plot.pdf}}
	\caption{Contours of integration for evaluating the Laplace inverse for the two cases namely $4\gamma>\omega(q)$ and $4\gamma<\omega(q)$. \new{$s_-$ in (a) is shown in blue as it lies on the second Riemann sheet and therefore does not contribute in the integral.} }
	\label{fig:cont1}
\end{figure}
It is not surprising that the analytic structure of the integrand in Eq.~(\ref{eq:Cxxfeq}) in the complex $s$ plane is important in determining the type of the behavior we observe. The evaluation of the  Laplace inverse in Eq.~(\ref{eq:Cxxfeq}) involves adding up the contribution due to the singularities of the function $\mathcal{F}(s,q)$.   While the details are present in the Appendix \ref{app:li}, we qualitatively discuss here the different contributions.

The integrand has  branch points at $b_\pm=\pm i\omega(q)$, and we take the branch cut to run along the imaginary axis between the two branch points.  There are also two poles  located at $s_\pm=\pm \sqrt{16\gamma^2-\omega^2(q)} $. The poles lie on the real axis for $4\gamma>\omega(q)$ and on the branch cut along imaginary axis for $4\gamma<\omega(q)$. The branch cut joins the two Riemann sheets and the contours of integration(see Fig.~\ref{fig:cont1}) for the Laplace inverse are taken on the principal Riemann sheet.

 The contribution due to the branch cut is ballistic, and is sensitive to the details of the spectrum of the Hamiltonian \new{i.e. the dispersion given by $\omega(q)$}.  When the poles lie on the imaginary axis, their contribution is ballistic akin to forward and backward propagating plane waves of frequency $\sqrt{\omega^2(q)-16\gamma^2}$.   However, as $q$ changes such that $\omega(q)$ approaches $4\gamma$ the poles sweep along the imaginary axis towards the origin and meet at the origin at $4\gamma=\omega(q)$. For $4\gamma>\omega(q)$ the poles  are  off the branch cut and  shift to the real axis but on different Riemann sheets joined by the branch cut. Therefore, one has to be careful about which pole contributes. $s_+$ shifts to the principal Riemann sheet, and $s_-$ shifts  to the second Riemann sheet. Thus, only $s_+$ contributes and gives a diffusive contribution. This is also evident from Eq.~(\ref{eq:Cxxsmalls}) where the contribution from the pole at $s=-2 q^2/\gamma$, which is just $s_+$ at small $q$ shifted by $4\gamma$, gives rise to the diffusion kernel.
\new{Physically, the contribution of the pole at $s_-$ and $s_+$ correspond to  evolution backward and forward in time, respectively. Because the diffusion equation (holding in  small $q$ or long-wavelength limit) has a meaning only evolving forward in time (going backwards in time any inhomogeneity would get sharper, instead of smoother, violating relaxation) only $s_+$ has to be kept. For the ballistic short-wavelength part at $4\gamma<\omega(q)$, where the dephasing scattering length $\sim 1/\gamma$ is larger than the wavelength $\sim 1/q$ and one deals with the wave equation. Therefore, both forward and backward in time evolution are physical and both contributions have to be included.}

  \begin{figure}
 	\centering
 	\subfigure{
 		\includegraphics[width=0.48\textwidth,page=3]{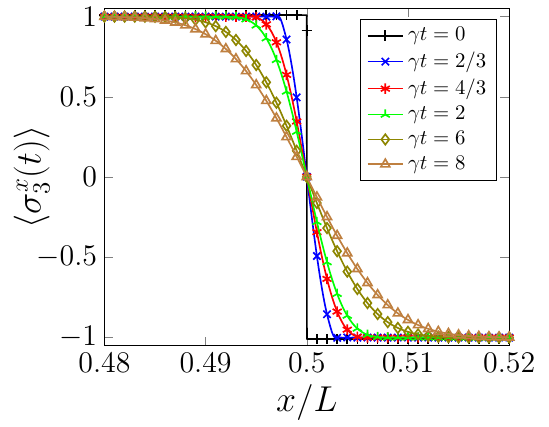}}
 	\caption{ Evolution of the density, near $x=L/2$, starting from a domain wall initial state for $L=10^5$ using Eq.~(\ref{eq:Cxxfeq}) with $\gamma=0.01,~J=1$ where the Laplace inverse has been taken numerically.}
 \label{fig:denevol1}
 \end{figure}
 \begin{figure}
 	\subfigure{
 		\includegraphics[width=0.47\textwidth,page=1]{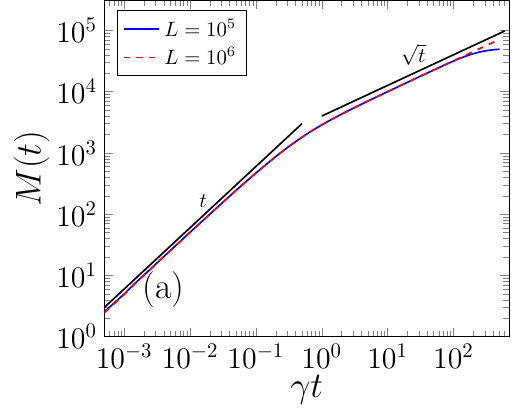}}
 	\subfigure{
 		\includegraphics[width=0.47\textwidth,page=2]{plot1.pdf}}
 	\caption{ (a) Variation of  the transferred magnetization $M(t)$ with time at $\gamma=0.01$. We see that  at short times, $\gamma t\ll1$, $M(t)\sim t$ and at long times, $\gamma t\gg1$, $M(t)\sim \sqrt{t}$. (b) Logarithmic derivative of the transferred magnetization with time. The red crosses, brown dots and the blue dashed line is $\beta(t)$ obtained by using Eq.~(\ref{eq:Cxxfeq}). The black solid line is the data from exact diagonalization of Eq.~(\ref{eq:correq}).}
 	\label{fig:denevol}
 \end{figure}

\subsection{Numerical Results}
 Eq.~(\ref{eq:Cxxfeq}) can be directly evaluated  numerically by summing over the allowed values of $q$ in place of the integral over $q$ and taking the Laplace inverse numerically  using standard approaches such as Talbot's method~\cite{dingfelder2015improved,cohen2007numerical}. A nice implementation of these methods can be found in $\textit{mpmath}$~\cite{mpmath} python library. \neww{ While for the XX model with dephasing we can use  Eq.~(\ref{eq:Cxxfeq}) directly to get the density} but in general one needs to multiply the transfer matrices and invert the product  to get $\mathcal{G}_{0,0}$ which is then Laplace inverted  for every allowed value of $q$. Therefore, assuming that the transfer matrices are of dimension $r$, the complexity of such a calculation scales as $L^2 r^6$. Given that $r$ is of the order of range of couplings which is usually small,  the complexity of this method is  better than obtaining the density at any time $t$ via exact diagonalization or iterations  of matrix Eq.~(\ref{eq:correq}) of size $\sim L^2$, which have a complexity of about $L^6$. Using the transfer matrix approach magnetization profiles for system sizes of the order of $10^6$  can be easily obtained which is much bigger than $L \approx 10^4$ feasible through diagonalization~\cite{varma2017fractality}.

 \new{
Fig.~\ref{fig:denevol1} shows the evolution of the magnetization density, $\langle \sigma_3^x\rangle=-C_{x,x}(t)$ for $L=10^5$ starting from a domain wall initial state using Eq.~(\ref{eq:Cxxfeq}). By domain wall initial state, we mean a product  state of Eq.~(\ref{eq:initstate}) where spins with $x<L/2$ are in the up state and spins with $x>L/2$ are in the down state. The slow down of the propagation of the front is clearly visible from Fig.~\ref{fig:denevol1} as $\gamma t$ crosses unit value.

  As the dynamics transitions from ballistic to diffusive, we also show the behavior of the transferred magnetization, defined as $M(t)=\sum_{x=(L-1)/2}^{L-1}\langle \sigma_3^x\rangle+L/2$,  with time. Panel (a) of Fig.~\ref{fig:denevol} shows the evolution of  $M(t)$ with $\gamma t$,  and we can clearly see the crossover  from $M(t)\sim t$ to $M(t)\sim \sqrt{t}$ indicating the expected change from ballistic to diffusive behavior. In order to study the change in the slope of the transferred magnetization at different $L$ and $\gamma$, we look at its logarithmic derivative, 
  \begin{equation}
 \beta(t)=\frac{d}{d\log t}\log[M(t)].
  \end{equation}

	The plot for $\beta(t)$ is shown in panel (b) of Fig.~\ref{fig:denevol}. We find that $\beta(t)$ shows oscillations which decay with time, see the brown dots $(L=10^5,\gamma=0.01)$ and red crosses ($L=1600$, $\gamma=0.01$) in Fig.~\ref{fig:denevol}b.  The oscillations arise as at small times the Green's function is given by  Bessel functions, and their amplitude goes down as a power-law (nearly $~1/t$). The bare time scale for the oscillations   is given by $t\sim 1/J$, which in the scaled time, $\gamma t$, goes as $\gamma/J$.  Therefore, as $\gamma$ is lowered the oscillations shift towards smaller and smaller values of $\gamma t$. We can see this in the blue dashed curve ($L=10^5$, $\gamma=0.001$) where the oscillations have already become too small to be captured by Talbot's numerical Laplace inverse, and one gets a monotonically decreasing curve from ballistic behavior, $\beta=1$, to diffusive behavior, $\beta=0.5$. Talbot's method fails to capture the oscillations as the oscillation  strength becomes small because of  its  limitations with oscillatory time domain functions. We confirm this by showing the data from exact diagonalization (ED), black solid line, for $L=1600$ and $\gamma=0.01$, where the oscillation show a regular decay while as there is sudden disappearance of oscillations near \neww{$\gamma t\approx 10^{-1}$} from the corresponding plot \neww{(red crosses)} obtained using the numerical Laplace inverse. 
	
	Note that  initially the brown dots overlap with the red crosses, as the initial dynamics will not show any finite size effects. However, for $  t\sim L/4$  for the red crosses, they start to sharply go below $\beta=0.5$ as the finite size effect kicks in, on the other hand the brown dots saturate to $\beta=0.5$. The factor of $1/4$ in the time scale for the finite size effects arises due to periodic boundaries as the two fronts in the domain wall at $x=L/2$ and $x=L$ move towards each other and meet when $t\sim L/4$.  }
 
 \subsection{Thermodynamic limit of the Off-diagonal Elements}
Let us now consider the off-diagonal elements  of the correlation matrix. Their solution in the limit $L\rightarrow\infty$ follows from using Eq.~(\ref{eq:G00inf}) in  Eq.~(\ref{eq:gl0}). Doing so and replacing $\sin(\alpha l)$ by $(e^{i\alpha l}-e^{-i\alpha l})/(2i)$ we get,
\begin{align}
	\mathcal{G}_{l,0}=\frac{e^{i \alpha l}}{2\omega\sin \alpha}\bigg(1+\frac{i\omega e^{-i\alpha}-s}{\sqrt{\tilde s^2+\omega^2}-4\gamma}\bigg),
\end{align}
where $\tilde s=s+4\gamma$ and we have denoted $\omega(q)$ as just $\omega$. We use Eq.~(\ref{eq:arcos}) to write the above equation as,
\begin{align}
 \mathcal{G}_{l,0}=\frac{i^l \omega^l(1+(4\gamma+\sqrt{\tilde s^2+\omega^2})(\sqrt{\tilde s^2+\omega^2}-4\gamma)^{-1})}{2\sqrt{\tilde s^2+\omega^2}\left(\tilde s+\sqrt{\tilde s^2+\omega^2}\right)^l}.
\end{align}
The solution for the correlators is obtained  by substituting this equation into Eq.~(\ref{eq:cltotildeg}). 

\begin{figure}
	\centering
	\subfigure{
		\includegraphics[width=0.47\textwidth,page=1]{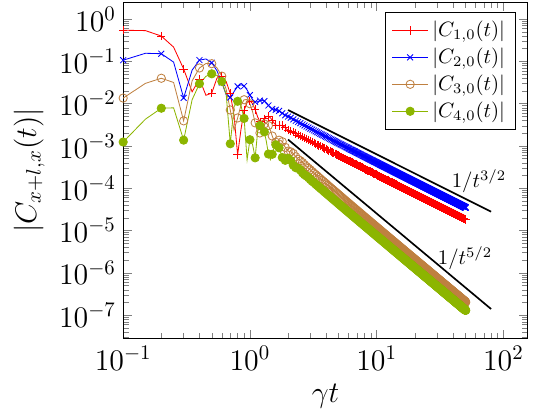}}
	\caption{Evolution of the off-diagonal elements with $\gamma=0.5$ and $L=200$. We see an agreement with the asymptotic behavior predicted by Eq.~(\ref{eq:Cxylt}).   }
	\label{fig:offd}
  \end{figure}

The asymptotic behaviors for long and short times  is  obtained  by looking at the small $s\sim \omega(q)<4\gamma$ and at the  large $s\sim \omega(q)>4\gamma$ behavior of $\mathcal{G}_{l,0}$, respectively. We once again choose the initial  condition to be $c_x=\delta_{x,0}$. The small $s$ behavior is given by,
\begin{equation}
\mathcal{G}_{l,0}^{b}=\frac{(i\omega)^l}{\sqrt{\tilde s^2+\omega^2}(\tilde s+\sqrt{\tilde s^2+\omega^2})^l},	
\end{equation}
where we have retained $4\gamma$ in $\tilde s$ as it  gives the overall decay of $e^{-4\gamma t}$ when substituted into Eq.~(\ref{eq:cltotildeg}) as follows,
\begin{align}
	&C_{x+l,x}\notag\\&= e^{-4\gamma t}\int_0^{2\pi} \frac{dq}{2\pi}   \mathcal{L}^{-1}\bigg[\frac{(-1)^le^{i q (x-l/2)}\omega^l(q)}{\sqrt{s^2+\omega^2(q)}\big( s+\sqrt{ s^2+\omega^2(q)}\big)^l}\bigg]. 
\end{align}
Similarly, the small $s$ limit of $\mathcal{G}_{l,0}$ gives,
\begin{equation}
	\mathcal{G}_{l,0}^d\approx\left(\frac{i  }{2\gamma}\right)^l\frac{q^l}{s+ 2J^2 q^2/\gamma}
\end{equation}
which when substituted into Eq.~(\ref{eq:cltotildeg}) leads to the following expression for the off-diagonal correlators in the long time limit,
\begin{align}
	C_{x+l,x}(t)\approx \left(\frac{i}{2\gamma}\right)^l \frac{\partial^l}{\partial x^l} \frac{e^{-\frac{(x-l/2)^2}{8J^2 t/\gamma}}}{\sqrt{8\pi J^2 t/\gamma}}.\label{eq:Cxylt}
\end{align}
Therefore, for $|x-l/2|<8 J^2t/\gamma$, the leading behavior of  $|C_{x+l,x}|\sim\frac{1}{(\gamma t)^{k+1/2}}$ for $l=2k-1$ and $l=2k$, where $k=1,2,3,...$. In Fig.~\ref{fig:offd}, we show that agreement of these behaviors with the exact numerical computation.

\section{Conclusion}
\label{sec:concl}
In conclusion, we demonstrated the use of transfer matrix approach to solve for the two point correlators of an XX spin chain with homogeneous dephasing. We showed that this approach leads to a simple solution \neww{for} all the elements of the correlation matrix in terms of products  of $2\times2$ transfer matrices. From the solution, the thermodynamic limit follows directly due to the explicit presence of $L$ dependence in the products. Our main result in Eq.~(\ref{eq:Cxxfeq}) provides a simple expression for the evolution of the magnetization density at any time, $t$ and   from this expression the asymptotic behaviors  at short and long times of the magnetization density become readily apparent. We also looked at the asymptotic behaviors of the off-diagonal elements and found that at long times they can be expressed as higher order spatial derivatives of the diagonal components. The latter gives that the long time behavior of the $l^{th}$ off-diagonal elements is given by $1/t^{\lceil l/2 \rceil+1/2}$.

While we only looked at a rather simple model with local dephasing, this approach can in principle be utilized to solve for the two point correlators  of other quadratic Hamiltonians subject to Hermitian  Lindblad operators that give a closed set of linear equations for the correlators. Clearly, for complicated Lindblad operators it may not be possible to get as simple  expressions as  Eq.~(\ref{eq:Cxxfeq}). \new{However, we can always obtain the transfer matrix equation of the type in Eq.~(\ref{eq:g0gN}) which, as we demonstrated, can be used numerically to determine the evolution of the  density. The complexity of the  transfer matrix method with transfer matrices of dimension $r$ is of the order of $L^2r^6$ as opposed to exact diagonalization or iterations of Eq.~(\ref{eq:correq}) which scales as $L^6$. Therefore, one is able to access large system sizes as long as the transfer matrices are of small dimension.} The dimensions of the transfer matrix  depend on the range of the couplings present in the system, and therefore will be small as long as the system has just nearest neighbor or next-to-nearest neighbor coupling. For example, the non-local dephasing models~\cite{vznidarivc2024superdiffusive,wang2024}  with three site dephasing operators has  a $4\times 4$ transfer matrix. 

{\em Note:} After completion of this work Ref.~\onlinecite{ishiyama2025exactdensityprofiletightbinding} appeared, which obtains compact expressions for the density profiles for XX spin chain with dephasing via the Bethe ansatz method. Our approach is different and could be used also in other models not solvable by Bethe ansatz.

\begin{acknowledgements}
	We acknowledge the support of Grant No.~J1-4385  from the Slovenian Research Agency.
 \end{acknowledgements}
\appendix
\section{Even-Odd structure of the correlation matrix}
\label{app:symm}
Here we show  that if the initial correlation matrix, $C_{x,y}(0)$, is such that its even off-diagonal elements are real and odd off-diagonal elements are imaginary then it remains so for all the times.  Assuming that it is true at time $t_n$, then at the next step
\begin{align}
  &C_{x,y}^{im}(t_{n+1})=C_{x,y}^{im}(t_{n})+\notag \bigg[2J(C_{x-1,y}^{re}(t_n)+C_{x+1,y}^{re}(t_n))\\&-4\gamma C^{im}_{x,y}(t_n)+2J(C_{x,y-1}^{re}(t_n)+C_{x,y+1}^{re}(t_n))\bigg](t_{n+1}-t_n),
\end{align}
where the superscripts $re$ and $im$ denote real and imaginary parts of $C_{xy}(t)$, respectively. Clearly, if $x-y$ is even then $C_{x,y}^{im}(t_{n+1})=0$ as it involves itself and only the real parts of its neighboring off-diagonals at time $t_n$ which by assumption are zero. Using the same argument $C_{x,y}^{re}(t_{n+1})=0$ if $x-y$ is odd as it involves only the imaginary parts of the neighboring off-diagonals at time $t_n$. Therefore, at any time $t$, the even off-diagonals will remain real while as the odd off-diagonals will remain imaginary.

\section{Derivation of the Transfer Matrix}
\label{app:tfm}
Let us consider the first column of equations from the identity $(s-A)\mathcal{G}(s)=I$ we have,
\begin{align}
	&s \mathcal{G}_{0,0}-i\omega(q_n)\mathcal{G}_{1,0} = 1,\label{aeq:ideq1}\\
	&-i  \frac{\omega(q_n)}{2} \mathcal{G}_{i-1,0}+(s+4\gamma) \mathcal{G}_{i,0}\notag\\	
	&~~~~~~~~~~~~~~~~~~-i\frac{\omega(q_n)}{2} \mathcal{G}_{i+1,0}=0,~0<i<L-1,\label{aeq:ideq2}\\
	&-i  \frac{\omega(q_n)}{2} \mathcal{G}_{L-2,0}+(s+4\gamma) \mathcal{G}_{L-1,0}-i \frac{\omega(q_n)}{2} \mathcal{G}_{0,0}=0.\label{aeq:ideq3}
\end{align}
We can rewrite Eq.~(\ref{aeq:ideq1}) as,
\begin{equation}
	\begin{pmatrix}
		1\\\mathcal{G}_{00}
	\end{pmatrix}=\begin{pmatrix}s & -i\omega(q_n)\\
		1 & 0
	\end{pmatrix}\begin{pmatrix}
		\mathcal{G}_{00}\\\mathcal{G}_{10}
	\end{pmatrix}\label{aeq:te1}
\end{equation}
We now use Eq.~(\ref{aeq:ideq2}) for $i=1$ to write $ \begin{pmatrix}
	\mathcal{G}_{00}\\\mathcal{G}_{10}
\end{pmatrix}$  in terms of $\begin{pmatrix}
	\mathcal{G}_{10}\\\mathcal{G}_{20}
\end{pmatrix}$ in Eq.~(\ref{aeq:te1}) and obtain the following,
\begin{equation}
	\begin{pmatrix}
		1\\\mathcal{G}_{00}
	\end{pmatrix}=\begin{pmatrix}s & -i\omega(q_n)\\
		1 & 0
	\end{pmatrix}\begin{pmatrix}
		\frac{2(s + 4\gamma)}{i\omega(q_n) } & -1 \\
		1 & 0
	\end{pmatrix}\begin{pmatrix}
		\mathcal{G}_{10}\\\mathcal{G}_{20}
	\end{pmatrix}.
\end{equation}

Carrying out the iterations similarly using Eq.~(\ref{aeq:ideq2}) for $i=2,3,..,L-2$  and then finally using Eq.~(\ref{aeq:ideq3}), we obtain the following,
\begin{align}
	\begin{pmatrix}
		1\\\mathcal{G}_{00}
	\end{pmatrix}&=	\begin{pmatrix}s & -i\omega(q_n)\\
		1 & 0
	\end{pmatrix}
	\begin{pmatrix}
		-2i u & -1 \\
		1 & 0
	\end{pmatrix}^{L-1}\begin{pmatrix}
		\mathcal{G}_{L-1,0}\\\mathcal{G}_{00}
	\end{pmatrix},\\
	\begin{pmatrix}
		1\\\mathcal{G}_{00}
	\end{pmatrix}&=T_0 T^{L-1}\begin{pmatrix}
		\mathcal{G}_{L-1,0}\\\mathcal{G}_{00}
	\end{pmatrix},\label{aeq:g0gN}
\end{align}
where $u=(s+4\gamma)/\omega(q_n)$. $T^{L-1}$ can be simply computed and is given by,
\begin{equation}
	T^{L-1}=\frac{1}{\sin(\alpha)}\begin{pmatrix}
		\sin(\alpha L) & -\sin(\alpha(L-1)) \\
		\sin(\alpha(L-1)) & -\sin(\alpha(L-2))
	\end{pmatrix},
\end{equation}
where $\alpha=\arccos[-iu].$

\section{Laplace Inverse}
\label{app:li}

 In this section we compute  the Laplace inverse in Eq.~(\ref{eq:Cxxfeq}). We recall the definition of Laplace inverse from Eq.~(\ref{eq:lapinv}),
\begin{align}
	I(t)&= \mathcal{L}^{-1}\Bigg[\frac{1}{\sqrt{s^2+\omega^2(q)}-4\gamma}\Bigg]\\&=\frac{1}{2\pi i }\int_{\eta-i\infty}^{\eta+i\infty}dse^{st}\frac{1}{\sqrt{s^2+\omega^2(q)}-4\gamma},
\end{align}
where $\eta$ is taken to be bigger than real parts of all the singularities of $(\sqrt{s^2+\omega^2(q)}-4\gamma)^{-1}$.  This integral can be computed using a Bromwich contour where the analytic structure of the integrand in the complex $s$ plane becomes important. As stated in the main text, the integrand has  a branch cut  which we choose  to run between the two branch points $b_\pm=\pm i\omega(q)$~(See Fig~(\ref{fig:cont1})). The integrand also has poles located at $s_\pm=\pm \sqrt{4\gamma^2-\omega^2(q)} $. The poles lie on the real axis for $4\gamma>\omega(q)$ and on the imaginary axis for $4\gamma<\omega(q)$.  Since $\omega(q)=8J\sin(q/2)$ is  $q$ dependent, the two cases of real and imaginary poles needs to be considered separately.
	
	\textit{Real Poles $(4\gamma>\omega(q))$:} In this case, the poles lie on the real axis at $s=\pm s_0$ where $s_0=\sqrt{|4\gamma^2-\omega^2(q)|}$ and therefore we consider the contour shown in  Fig.~\ref{fig:cont1}a on the principal Riemann sheet. The branch cut contributes to the integral as usual, however, only one of the two poles, $s_+$, contributes. The reason being that the other pole $s_-$ lies on the second Riemann sheet and therefore lies outside the contour. \new{Evaluating the integral gives the following,
	\begin{align}
		I_r(t,q)= 4\gamma & \frac{e^{t\sqrt{4\gamma^2-\omega^2}}}{\sqrt{4\gamma^2-\omega^2}}\notag\\&+\frac{2}{\pi}\int_0^\omega ds \cos(st)\frac{\sqrt{\omega^2-s^2}}{\omega^2-16\gamma^2-s^2},\label{eq:Id}
	\end{align}
where the first term comes from the residue at $s=s_+$ and the second term, the integral, comes from the branch cut. }

\textit{ Imaginary Poles $(4\gamma<\omega(q))$:} The poles now lie on the branch cut at $s=\pm i s_0$, so we chose the contour shown in Fig.~\ref{fig:cont1}b. The integral over the branch cut now avoids the two poles and thus the integration will  be the sum of the residues due to the two poles together with the principal value of the integral over the branch cut. \new{ One obtains the following,
	\begin{align}
		I_i(t,q)= 4\gamma &\frac{\sin(t\sqrt{\omega^2-4\gamma^2})}{\sqrt{\omega^2-4\gamma^2}}\notag\\&+\frac{2}{\pi}\dashint_0^\omega ds \cos(st)\frac{\sqrt{\omega^2-s^2}}{\omega^2-16\gamma^2-s^2},\label{eq:Ib}
	\end{align}
where the first term is the sum of the contributions due to the two poles on the branch cut and the second term, the integral, is the contribution from the rest of the branch cut. The dash sign on the integral denotes the principal value of the integral which is taken by avoiding  the singularity at $s=\sqrt{\omega^2-16\gamma^2}.$}


Using these in Eq.~(\ref{eq:Cxxfeq}) we get,
\begin{align}
	C_{xx}(t)=&\int_{|\omega(q)|<4\gamma} dqe^{i q x} c(q)e^{-4\gamma t} I_r(t,q)\notag\\&+\int_{|\omega(q)|>4\gamma} dq e^{i q x} c(q)e^{-4\gamma t} I_i(t,q),
\end{align}
where  $I_r(t,q)$, $I_i(t,q)$ are given  by  Eq.~(\ref{eq:Id}) and Eq.~(\ref{eq:Ib}), respectively.

\new{
 One can quickly see that in the long time limit the exponential damping, $e^{-4\gamma t}$, kills off all the terms except the long wavelength contribution from the pole on the real axis i.e. the first term in $I_r(t,q)$ at small $q$. Therefore, dropping all other terms and expanding the first term in $I_r(t,q)$ around $q=0$, we recover the diffusion kernel.  To get the short time behavior requires some algebra. To that end, rewriting the integrals in $I_r(t,q)$ and $I_i(t,q)$ in terms of the variable $s=\omega\cos\theta$, the resulting expressions for $C_{x,x}(t)$ become the same as in Ref.~\onlinecite{eisler2011crossover} upto a re-parametrization. One can then do similar analysis as in Ref.~\onlinecite{eisler2011crossover} to recover the result in Eq.~(\ref{eq:Cxxlarges}).}

\bibliography{biblio_1.bib}

\begin{thebibliography}{47}%
\makeatletter
\providecommand \@ifxundefined [1]{%
 \@ifx{#1\undefined}
}%
\providecommand \@ifnum [1]{%
 \ifnum #1\expandafter \@firstoftwo
 \else \expandafter \@secondoftwo
 \fi
}%
\providecommand \@ifx [1]{%
 \ifx #1\expandafter \@firstoftwo
 \else \expandafter \@secondoftwo
 \fi
}%
\providecommand \natexlab [1]{#1}%
\providecommand \enquote  [1]{``#1''}%
\providecommand \bibnamefont  [1]{#1}%
\providecommand \bibfnamefont [1]{#1}%
\providecommand \citenamefont [1]{#1}%
\providecommand \href@noop [0]{\@secondoftwo}%
\providecommand \href [0]{\begingroup \@sanitize@url \@href}%
\providecommand \@href[1]{\@@startlink{#1}\@@href}%
\providecommand \@@href[1]{\endgroup#1\@@endlink}%
\providecommand \@sanitize@url [0]{\catcode `\\12\catcode `\$12\catcode
  `\&12\catcode `\#12\catcode `\^12\catcode `\_12\catcode `\%12\relax}%
\providecommand \@@startlink[1]{}%
\providecommand \@@endlink[0]{}%
\providecommand \url  [0]{\begingroup\@sanitize@url \@url }%
\providecommand \@url [1]{\endgroup\@href {#1}{\urlprefix }}%
\providecommand \urlprefix  [0]{URL }%
\providecommand \Eprint [0]{\href }%
\providecommand \doibase [0]{http://dx.doi.org/}%
\providecommand \selectlanguage [0]{\@gobble}%
\providecommand \bibinfo  [0]{\@secondoftwo}%
\providecommand \bibfield  [0]{\@secondoftwo}%
\providecommand \translation [1]{[#1]}%
\providecommand \BibitemOpen [0]{}%
\providecommand \bibitemStop [0]{}%
\providecommand \bibitemNoStop [0]{.\EOS\space}%
\providecommand \EOS [0]{\spacefactor3000\relax}%
\providecommand \BibitemShut  [1]{\csname bibitem#1\endcsname}%
\let\auto@bib@innerbib\@empty
\bibitem [{\citenamefont {Yeh}(2012)}]{yeh2012applied}%
  \BibitemOpen
  \bibfield  {author} {\bibinfo {author} {\bibfnamefont {C.}~\bibnamefont
  {Yeh}},\ }\href@noop {} {\emph {\bibinfo {title} {Applied photonics}}}\
  (\bibinfo  {publisher} {Elsevier},\ \bibinfo {year} {2012})\BibitemShut
  {NoStop}%
\bibitem [{\citenamefont {Pendry}\ and\ \citenamefont
  {MacKinnon}(1992)}]{pendry1992calculation}%
  \BibitemOpen
  \bibfield  {author} {\bibinfo {author} {\bibfnamefont {J.~B.}\ \bibnamefont
  {Pendry}}\ and\ \bibinfo {author} {\bibfnamefont {A.}~\bibnamefont
  {MacKinnon}},\ }\bibfield  {title} {\enquote {\bibinfo {title} {Calculation
  of photon dispersion relations},}\ }\href@noop {} {\bibfield  {journal}
  {\bibinfo  {journal} {Physical Review Letters}\ }\textbf {\bibinfo {volume}
  {69}},\ \bibinfo {pages} {2772} (\bibinfo {year} {1992})}\BibitemShut
  {NoStop}%
\bibitem [{\citenamefont {Pendry}\ \emph {et~al.}(1992)\citenamefont {Pendry},
  \citenamefont {MacKinnon},\ and\ \citenamefont
  {Roberts}}]{pendry1992universality}%
  \BibitemOpen
  \bibfield  {author} {\bibinfo {author} {\bibfnamefont {J.~B.}\ \bibnamefont
  {Pendry}}, \bibinfo {author} {\bibfnamefont {A.}~\bibnamefont {MacKinnon}}, \
  and\ \bibinfo {author} {\bibfnamefont {P.~J.}\ \bibnamefont {Roberts}},\
  }\bibfield  {title} {\enquote {\bibinfo {title} {Universality classes and
  fluctuations in disordered systems},}\ }\href@noop {} {\bibfield  {journal}
  {\bibinfo  {journal} {Proceedings of the Royal Society of London. Series A:
  Mathematical and Physical Sciences}\ }\textbf {\bibinfo {volume} {437}},\
  \bibinfo {pages} {67--83} (\bibinfo {year} {1992})}\BibitemShut {NoStop}%
\bibitem [{\citenamefont {Mello}\ and\ \citenamefont
  {Kumar}(2004)}]{mello2004quantum}%
  \BibitemOpen
  \bibfield  {author} {\bibinfo {author} {\bibfnamefont {P.~A.}\ \bibnamefont
  {Mello}}\ and\ \bibinfo {author} {\bibfnamefont {N.}~\bibnamefont {Kumar}},\
  }\href@noop {} {\emph {\bibinfo {title} {Quantum Transport in Mesoscopic
  Systems: Complexity and Statistical Fluctuations. A Maximum Entropy
  Viewpoint}}}\ (\bibinfo  {publisher} {Oxford University Press},\ \bibinfo
  {year} {2004})\BibitemShut {NoStop}%
\bibitem [{\citenamefont {Beenakker}(1997)}]{beenakker1997random}%
  \BibitemOpen
  \bibfield  {author} {\bibinfo {author} {\bibfnamefont {C.~W.~J.}\
  \bibnamefont {Beenakker}},\ }\bibfield  {title} {\enquote {\bibinfo {title}
  {Random-matrix theory of quantum transport},}\ }\href@noop {} {\bibfield
  {journal} {\bibinfo  {journal} {Reviews of modern physics}\ }\textbf
  {\bibinfo {volume} {69}},\ \bibinfo {pages} {731} (\bibinfo {year}
  {1997})}\BibitemShut {NoStop}%
\bibitem [{\citenamefont {Kirkman}\ and\ \citenamefont
  {Pendry}(1984)}]{kirkman1984statistics}%
  \BibitemOpen
  \bibfield  {author} {\bibinfo {author} {\bibfnamefont {P.~D.}\ \bibnamefont
  {Kirkman}}\ and\ \bibinfo {author} {\bibfnamefont {J.~B.}\ \bibnamefont
  {Pendry}},\ }\bibfield  {title} {\enquote {\bibinfo {title} {The statistics
  of the conductance of one-dimensional disordered chains},}\ }\href@noop {}
  {\bibfield  {journal} {\bibinfo  {journal} {Journal of Physics C: Solid State
  Physics}\ }\textbf {\bibinfo {volume} {17}},\ \bibinfo {pages} {5707}
  (\bibinfo {year} {1984})}\BibitemShut {NoStop}%
\bibitem [{\citenamefont {Ando}(1989)}]{ando1989numerical}%
  \BibitemOpen
  \bibfield  {author} {\bibinfo {author} {\bibfnamefont {T.}~\bibnamefont
  {Ando}},\ }\bibfield  {title} {\enquote {\bibinfo {title} {Numerical study of
  symmetry effects on localization in two dimensions},}\ }\href@noop {}
  {\bibfield  {journal} {\bibinfo  {journal} {Physical Review B}\ }\textbf
  {\bibinfo {volume} {40}},\ \bibinfo {pages} {5325} (\bibinfo {year}
  {1989})}\BibitemShut {NoStop}%
\bibitem [{\citenamefont {Pendry}(1982)}]{pendry1982evolution}%
  \BibitemOpen
  \bibfield  {author} {\bibinfo {author} {\bibfnamefont {J.~B.}\ \bibnamefont
  {Pendry}},\ }\bibfield  {title} {\enquote {\bibinfo {title} {The evolution of
  waves in disordered media},}\ }\href@noop {} {\bibfield  {journal} {\bibinfo
  {journal} {Journal of Physics C: Solid State Physics}\ }\textbf {\bibinfo
  {volume} {15}},\ \bibinfo {pages} {3493} (\bibinfo {year}
  {1982})}\BibitemShut {NoStop}%
\bibitem [{\citenamefont {Pendry}(1984)}]{pendry1984transfer}%
  \BibitemOpen
  \bibfield  {author} {\bibinfo {author} {\bibfnamefont {J.~B.}\ \bibnamefont
  {Pendry}},\ }\bibfield  {title} {\enquote {\bibinfo {title} {A transfer
  matrix approach to localisation in 3d},}\ }\href@noop {} {\bibfield
  {journal} {\bibinfo  {journal} {Journal of Physics C: Solid State Physics}\
  }\textbf {\bibinfo {volume} {17}},\ \bibinfo {pages} {5317} (\bibinfo {year}
  {1984})}\BibitemShut {NoStop}%
\bibitem [{\citenamefont {Lee}\ and\ \citenamefont
  {Joannopoulos}(1981{\natexlab{a}})}]{Simple1981Lee1}%
  \BibitemOpen
  \bibfield  {author} {\bibinfo {author} {\bibfnamefont {D.~H.}\ \bibnamefont
  {Lee}}\ and\ \bibinfo {author} {\bibfnamefont {J.~D.}\ \bibnamefont
  {Joannopoulos}},\ }\bibfield  {title} {\enquote {\bibinfo {title} {Simple
  scheme for surface-band calculations. {I}},}\ }\href@noop {} {\bibfield
  {journal} {\bibinfo  {journal} {Phys. Rev. B}\ }\textbf {\bibinfo {volume}
  {23}},\ \bibinfo {pages} {4988--4996} (\bibinfo {year}
  {1981}{\natexlab{a}})}\BibitemShut {NoStop}%
\bibitem [{\citenamefont {Lee}\ and\ \citenamefont
  {Joannopoulos}(1981{\natexlab{b}})}]{Simple1981Lee2}%
  \BibitemOpen
  \bibfield  {author} {\bibinfo {author} {\bibfnamefont {D.~H.}\ \bibnamefont
  {Lee}}\ and\ \bibinfo {author} {\bibfnamefont {J.~D.}\ \bibnamefont
  {Joannopoulos}},\ }\bibfield  {title} {\enquote {\bibinfo {title} {Simple
  scheme for surface-band calculations. {II}. the {G}reen's function},}\
  }\href@noop {} {\bibfield  {journal} {\bibinfo  {journal} {Phys. Rev. B}\
  }\textbf {\bibinfo {volume} {23}},\ \bibinfo {pages} {4997--5004} (\bibinfo
  {year} {1981}{\natexlab{b}})}\BibitemShut {NoStop}%
\bibitem [{\citenamefont {Izergin}\ \emph {et~al.}(1992)\citenamefont
  {Izergin}, \citenamefont {Coker},\ and\ \citenamefont
  {Korepin}}]{izergin1992determinant}%
  \BibitemOpen
  \bibfield  {author} {\bibinfo {author} {\bibfnamefont {A.~G.}\ \bibnamefont
  {Izergin}}, \bibinfo {author} {\bibfnamefont {D.~A.}\ \bibnamefont {Coker}},
  \ and\ \bibinfo {author} {\bibfnamefont {V.~E.}\ \bibnamefont {Korepin}},\
  }\bibfield  {title} {\enquote {\bibinfo {title} {Determinant formula for the
  six-vertex model},}\ }\href@noop {} {\bibfield  {journal} {\bibinfo
  {journal} {Journal of Physics A: Mathematical and General}\ }\textbf
  {\bibinfo {volume} {25}},\ \bibinfo {pages} {4315} (\bibinfo {year}
  {1992})}\BibitemShut {NoStop}%
\bibitem [{\citenamefont {Schultz}\ \emph {et~al.}(1964)\citenamefont
  {Schultz}, \citenamefont {Mattis},\ and\ \citenamefont
  {Lieb}}]{schultz1964two}%
  \BibitemOpen
  \bibfield  {author} {\bibinfo {author} {\bibfnamefont {T.~D.}\ \bibnamefont
  {Schultz}}, \bibinfo {author} {\bibfnamefont {D.~C.}\ \bibnamefont {Mattis}},
  \ and\ \bibinfo {author} {\bibfnamefont {E.~H.}\ \bibnamefont {Lieb}},\
  }\bibfield  {title} {\enquote {\bibinfo {title} {Two-dimensional ising model
  as a soluble problem of many fermions},}\ }\href@noop {} {\bibfield
  {journal} {\bibinfo  {journal} {Reviews of Modern Physics}\ }\textbf
  {\bibinfo {volume} {36}},\ \bibinfo {pages} {856} (\bibinfo {year}
  {1964})}\BibitemShut {NoStop}%
\bibitem [{\citenamefont {Akaike}(1973)}]{akaike1973block}%
  \BibitemOpen
  \bibfield  {author} {\bibinfo {author} {\bibfnamefont {H.}~\bibnamefont
  {Akaike}},\ }\bibfield  {title} {\enquote {\bibinfo {title} {Block {T}oeplitz
  matrix inversion},}\ }\href@noop {} {\bibfield  {journal} {\bibinfo
  {journal} {SIAM Journal on Applied Mathematics}\ }\textbf {\bibinfo {volume}
  {24}},\ \bibinfo {pages} {234--241} (\bibinfo {year} {1973})}\BibitemShut
  {NoStop}%
\bibitem [{\citenamefont {Molinari}(1997)}]{molinari1997transfer}%
  \BibitemOpen
  \bibfield  {author} {\bibinfo {author} {\bibfnamefont {L.}~\bibnamefont
  {Molinari}},\ }\bibfield  {title} {\enquote {\bibinfo {title} {Transfer
  matrices and tridiagonal-block {H}amiltonians with periodic and scattering
  boundary conditions},}\ }\href@noop {} {\bibfield  {journal} {\bibinfo
  {journal} {J. Phys. A: Math. Gen.}\ }\textbf {\bibinfo {volume} {30}},\
  \bibinfo {pages} {983} (\bibinfo {year} {1997})}\BibitemShut {NoStop}%
\bibitem [{\citenamefont {Reuter}\ and\ \citenamefont
  {Hill}(2012)}]{reuter2012efficient}%
  \BibitemOpen
  \bibfield  {author} {\bibinfo {author} {\bibfnamefont {M.~G.}\ \bibnamefont
  {Reuter}}\ and\ \bibinfo {author} {\bibfnamefont {J.~C.}\ \bibnamefont
  {Hill}},\ }\bibfield  {title} {\enquote {\bibinfo {title} {An efficient,
  block-by-block algorithm for inverting a block tridiagonal, nearly block
  {T}oeplitz matrix},}\ }\href@noop {} {\bibfield  {journal} {\bibinfo
  {journal} {Computational Science \& Discovery}\ }\textbf {\bibinfo {volume}
  {5}},\ \bibinfo {pages} {014009} (\bibinfo {year} {2012})}\BibitemShut
  {NoStop}%
\bibitem [{\citenamefont {Reuter}\ \emph {et~al.}(2011)\citenamefont {Reuter},
  \citenamefont {Seideman},\ and\ \citenamefont {Ratner}}]{reuter2011probing}%
  \BibitemOpen
  \bibfield  {author} {\bibinfo {author} {\bibfnamefont {M.~G.}\ \bibnamefont
  {Reuter}}, \bibinfo {author} {\bibfnamefont {T.}~\bibnamefont {Seideman}}, \
  and\ \bibinfo {author} {\bibfnamefont {M.~A.}\ \bibnamefont {Ratner}},\
  }\bibfield  {title} {\enquote {\bibinfo {title} {Probing the surface-to-bulk
  transition: {A} closed-form constant-scaling algorithm for computing
  subsurface {G}reen functions},}\ }\href@noop {} {\bibfield  {journal}
  {\bibinfo  {journal} {Phys. Rev. B}\ }\textbf {\bibinfo {volume} {83}},\
  \bibinfo {pages} {085412} (\bibinfo {year} {2011})}\BibitemShut {NoStop}%
\bibitem [{\citenamefont {Dhar}\ and\ \citenamefont
  {Sen}(2006)}]{dhar2006nonequilibrium}%
  \BibitemOpen
  \bibfield  {author} {\bibinfo {author} {\bibfnamefont {A.}~\bibnamefont
  {Dhar}}\ and\ \bibinfo {author} {\bibfnamefont {D.}~\bibnamefont {Sen}},\
  }\bibfield  {title} {\enquote {\bibinfo {title} {Nonequilibrium {G}reen’s
  function formalism and the problem of bound states},}\ }\href@noop {}
  {\bibfield  {journal} {\bibinfo  {journal} {Phys. Rev. B}\ }\textbf {\bibinfo
  {volume} {73}},\ \bibinfo {pages} {085119} (\bibinfo {year}
  {2006})}\BibitemShut {NoStop}%
\bibitem [{\citenamefont {Dhar}\ and\ \citenamefont
  {Roy}(2006)}]{dhar2006heat}%
  \BibitemOpen
  \bibfield  {author} {\bibinfo {author} {\bibfnamefont {A.}~\bibnamefont
  {Dhar}}\ and\ \bibinfo {author} {\bibfnamefont {D.}~\bibnamefont {Roy}},\
  }\bibfield  {title} {\enquote {\bibinfo {title} {Heat transport in harmonic
  lattices},}\ }\href@noop {} {\bibfield  {journal} {\bibinfo  {journal}
  {Journal of Statistical Physics}\ }\textbf {\bibinfo {volume} {125}},\
  \bibinfo {pages} {801--820} (\bibinfo {year} {2006})}\BibitemShut {NoStop}%
\bibitem [{\citenamefont {{\v{Z}}nidari{\v{c}}}(2010)}]{vznidarivc2010exact}%
  \BibitemOpen
  \bibfield  {author} {\bibinfo {author} {\bibfnamefont {M.}~\bibnamefont
  {{\v{Z}}nidari{\v{c}}}},\ }\bibfield  {title} {\enquote {\bibinfo {title}
  {Exact solution for a diffusive nonequilibrium steady state of an open
  quantum chain},}\ }\href@noop {} {\bibfield  {journal} {\bibinfo  {journal}
  {Journal of Statistical Mechanics: Theory and Experiment}\ }\textbf {\bibinfo
  {volume} {2010}},\ \bibinfo {pages} {L05002} (\bibinfo {year}
  {2010})}\BibitemShut {NoStop}%
\bibitem [{\citenamefont {Žunkovič}(2014)}]{zunk_2014}%
  \BibitemOpen
  \bibfield  {author} {\bibinfo {author} {\bibfnamefont {B.}~\bibnamefont
  {Žunkovič}},\ }\bibfield  {title} {\enquote {\bibinfo {title} {Closed
  hierarchy of correlations in markovian open quantum systems},}\ }\href@noop
  {} {\bibfield  {journal} {\bibinfo  {journal} {New Journal of Physics}\
  }\textbf {\bibinfo {volume} {16}},\ \bibinfo {pages} {013042} (\bibinfo
  {year} {2014})}\BibitemShut {NoStop}%
\bibitem [{\citenamefont {Barthel}\ and\ \citenamefont
  {Zhang}(2022)}]{barthel2022solving}%
  \BibitemOpen
  \bibfield  {author} {\bibinfo {author} {\bibfnamefont {T.}~\bibnamefont
  {Barthel}}\ and\ \bibinfo {author} {\bibfnamefont {Y.}~\bibnamefont
  {Zhang}},\ }\bibfield  {title} {\enquote {\bibinfo {title} {Solving
  quasi-free and quadratic lindblad master equations for open fermionic and
  bosonic systems},}\ }\href@noop {} {\bibfield  {journal} {\bibinfo  {journal}
  {Journal of Statistical Mechanics: Theory and Experiment}\ }\textbf {\bibinfo
  {volume} {2022}},\ \bibinfo {pages} {113101} (\bibinfo {year}
  {2022})}\BibitemShut {NoStop}%
\bibitem [{\citenamefont {Wang}\ \emph {et~al.}(2024)\citenamefont {Wang},
  \citenamefont {Fang},\ and\ \citenamefont {Ren}}]{wang2024}%
  \BibitemOpen
  \bibfield  {author} {\bibinfo {author} {\bibfnamefont {Y.~P.}\ \bibnamefont
  {Wang}}, \bibinfo {author} {\bibfnamefont {C.}~\bibnamefont {Fang}}, \ and\
  \bibinfo {author} {\bibfnamefont {J.}~\bibnamefont {Ren}},\ }\bibfield
  {title} {\enquote {\bibinfo {title} {{Superdiffusive transport in
  quasi-particle dephasing models}},}\ }\href@noop {} {\bibfield  {journal}
  {\bibinfo  {journal} {SciPost Phys.}\ }\textbf {\bibinfo {volume} {17}},\
  \bibinfo {pages} {150} (\bibinfo {year} {2024})}\BibitemShut {NoStop}%
\bibitem [{\citenamefont
  {{\v{Z}}nidari{\v{c}}}(2011)}]{vznidarivc2011solvable}%
  \BibitemOpen
  \bibfield  {author} {\bibinfo {author} {\bibfnamefont {M.}~\bibnamefont
  {{\v{Z}}nidari{\v{c}}}},\ }\bibfield  {title} {\enquote {\bibinfo {title}
  {Solvable quantum nonequilibrium model exhibiting a phase transition and a
  matrix product representation},}\ }\href@noop {} {\bibfield  {journal}
  {\bibinfo  {journal} {Physical Review E—Statistical, Nonlinear, and Soft
  Matter Physics}\ }\textbf {\bibinfo {volume} {83}},\ \bibinfo {pages}
  {011108} (\bibinfo {year} {2011})}\BibitemShut {NoStop}%
\bibitem [{\citenamefont
  {{\v{Z}}nidari{\v{c}}}(2024)}]{vznidarivc2024superdiffusive}%
  \BibitemOpen
  \bibfield  {author} {\bibinfo {author} {\bibfnamefont {M.}~\bibnamefont
  {{\v{Z}}nidari{\v{c}}}},\ }\bibfield  {title} {\enquote {\bibinfo {title}
  {Superdiffusive magnetization transport in the {XX} spin chain with nonlocal
  dephasing},}\ }\href@noop {} {\bibfield  {journal} {\bibinfo  {journal}
  {Physical Review B}\ }\textbf {\bibinfo {volume} {109}},\ \bibinfo {pages}
  {075105} (\bibinfo {year} {2024})}\BibitemShut {NoStop}%
\bibitem [{\citenamefont {Eisler}(2011)}]{eisler2011crossover}%
  \BibitemOpen
  \bibfield  {author} {\bibinfo {author} {\bibfnamefont {V.}~\bibnamefont
  {Eisler}},\ }\bibfield  {title} {\enquote {\bibinfo {title} {Crossover
  between ballistic and diffusive transport: the quantum exclusion process},}\
  }\href@noop {} {\bibfield  {journal} {\bibinfo  {journal} {Journal of
  Statistical Mechanics: Theory and Experiment}\ }\textbf {\bibinfo {volume}
  {2011}},\ \bibinfo {pages} {P06007} (\bibinfo {year} {2011})}\BibitemShut
  {NoStop}%
\bibitem [{\citenamefont {Temme}\ \emph {et~al.}(2012)\citenamefont {Temme},
  \citenamefont {Wolf},\ and\ \citenamefont
  {Verstraete}}]{temme2012stochastic}%
  \BibitemOpen
  \bibfield  {author} {\bibinfo {author} {\bibfnamefont {K.}~\bibnamefont
  {Temme}}, \bibinfo {author} {\bibfnamefont {M.~M.}\ \bibnamefont {Wolf}}, \
  and\ \bibinfo {author} {\bibfnamefont {F.}~\bibnamefont {Verstraete}},\
  }\bibfield  {title} {\enquote {\bibinfo {title} {Stochastic exclusion
  processes versus coherent transport},}\ }\href@noop {} {\bibfield  {journal}
  {\bibinfo  {journal} {New Journal of Physics}\ }\textbf {\bibinfo {volume}
  {14}},\ \bibinfo {pages} {075004} (\bibinfo {year} {2012})}\BibitemShut
  {NoStop}%
\bibitem [{\citenamefont {Medvedyeva}\ \emph {et~al.}(2016)\citenamefont
  {Medvedyeva}, \citenamefont {Essler},\ and\ \citenamefont
  {Prosen}}]{medvedyeva2016exact}%
  \BibitemOpen
  \bibfield  {author} {\bibinfo {author} {\bibfnamefont {M.~V.}\ \bibnamefont
  {Medvedyeva}}, \bibinfo {author} {\bibfnamefont {F.~H.~L.}\ \bibnamefont
  {Essler}}, \ and\ \bibinfo {author} {\bibfnamefont {T.}~\bibnamefont
  {Prosen}},\ }\bibfield  {title} {\enquote {\bibinfo {title} {Exact {B}ethe
  ansatz spectrum of a tight-binding chain with dephasing noise},}\ }\href@noop
  {} {\bibfield  {journal} {\bibinfo  {journal} {Physical review letters}\
  }\textbf {\bibinfo {volume} {117}},\ \bibinfo {pages} {137202} (\bibinfo
  {year} {2016})}\BibitemShut {NoStop}%
\bibitem [{\citenamefont {Teretenkov}\ and\ \citenamefont
  {Lychkovskiy}(2024)}]{Alex2024}%
  \BibitemOpen
  \bibfield  {author} {\bibinfo {author} {\bibfnamefont {A.}~\bibnamefont
  {Teretenkov}}\ and\ \bibinfo {author} {\bibfnamefont {O.}~\bibnamefont
  {Lychkovskiy}},\ }\bibfield  {title} {\enquote {\bibinfo {title} {Exact
  dynamics of quantum dissipative {XX} models: Wannier-stark localization in
  the fragmented operator space},}\ }\href@noop {} {\bibfield  {journal}
  {\bibinfo  {journal} {Phys. Rev. B}\ }\textbf {\bibinfo {volume} {109}},\
  \bibinfo {pages} {L140302} (\bibinfo {year} {2024})}\BibitemShut {NoStop}%
\bibitem [{\citenamefont {Silva}\ \emph {et~al.}(2023)\citenamefont {Silva},
  \citenamefont {Landi},\ and\ \citenamefont {Pereira}}]{silva2023nontrivial}%
  \BibitemOpen
  \bibfield  {author} {\bibinfo {author} {\bibfnamefont {S.~H.~S.}\
  \bibnamefont {Silva}}, \bibinfo {author} {\bibfnamefont {G.~T.}\ \bibnamefont
  {Landi}}, \ and\ \bibinfo {author} {\bibfnamefont {E.}~\bibnamefont
  {Pereira}},\ }\bibfield  {title} {\enquote {\bibinfo {title} {Nontrivial
  effect of dephasing: Enhancement of rectification of spin current in graded
  {XX} chains},}\ }\href@noop {} {\bibfield  {journal} {\bibinfo  {journal}
  {Physical Review E}\ }\textbf {\bibinfo {volume} {107}},\ \bibinfo {pages}
  {054123} (\bibinfo {year} {2023})}\BibitemShut {NoStop}%
\bibitem [{\citenamefont {Cao}\ \emph {et~al.}(2019)\citenamefont {Cao},
  \citenamefont {Tilloy},\ and\ \citenamefont {De~Luca}}]{caoSciPost19}%
  \BibitemOpen
  \bibfield  {author} {\bibinfo {author} {\bibfnamefont {X.}~\bibnamefont
  {Cao}}, \bibinfo {author} {\bibfnamefont {A.}~\bibnamefont {Tilloy}}, \ and\
  \bibinfo {author} {\bibfnamefont {A.}~\bibnamefont {De~Luca}},\ }\bibfield
  {title} {\enquote {\bibinfo {title} {Entanglement in a fermion chain under
  continuous monitoring},}\ }\href@noop {} {\bibfield  {journal} {\bibinfo
  {journal} {SciPost Physics}\ }\textbf {\bibinfo {volume} {7}},\ \bibinfo
  {pages} {024} (\bibinfo {year} {2019})}\BibitemShut {NoStop}%
\bibitem [{\citenamefont {Turkeshi}\ and\ \citenamefont
  {Schir{\'o}}(2021)}]{turkeshiPRB21}%
  \BibitemOpen
  \bibfield  {author} {\bibinfo {author} {\bibfnamefont {X.}~\bibnamefont
  {Turkeshi}}\ and\ \bibinfo {author} {\bibfnamefont {M.}~\bibnamefont
  {Schir{\'o}}},\ }\bibfield  {title} {\enquote {\bibinfo {title} {Diffusion
  and thermalization in a boundary-driven dephasing model},}\ }\href@noop {}
  {\bibfield  {journal} {\bibinfo  {journal} {Physical Review B}\ }\textbf
  {\bibinfo {volume} {104}},\ \bibinfo {pages} {144301} (\bibinfo {year}
  {2021})}\BibitemShut {NoStop}%
\bibitem [{\citenamefont
  {{\v{Z}}nidari{\v{c}}}(2015)}]{vznidarivc2015relaxation}%
  \BibitemOpen
  \bibfield  {author} {\bibinfo {author} {\bibfnamefont {M.}~\bibnamefont
  {{\v{Z}}nidari{\v{c}}}},\ }\bibfield  {title} {\enquote {\bibinfo {title}
  {Relaxation times of dissipative many-body quantum systems},}\ }\href@noop {}
  {\bibfield  {journal} {\bibinfo  {journal} {Physical Review E}\ }\textbf
  {\bibinfo {volume} {92}},\ \bibinfo {pages} {042143} (\bibinfo {year}
  {2015})}\BibitemShut {NoStop}%
\bibitem [{\citenamefont {Carollo}\ \emph {et~al.}(2017)\citenamefont
  {Carollo}, \citenamefont {Garrahan}, \citenamefont {Lesanovsky},\ and\
  \citenamefont {P{\'e}rez-Espigares}}]{carolloPRE17}%
  \BibitemOpen
  \bibfield  {author} {\bibinfo {author} {\bibfnamefont {F.}~\bibnamefont
  {Carollo}}, \bibinfo {author} {\bibfnamefont {J.~P.}\ \bibnamefont
  {Garrahan}}, \bibinfo {author} {\bibfnamefont {I.}~\bibnamefont
  {Lesanovsky}}, \ and\ \bibinfo {author} {\bibfnamefont {C.}~\bibnamefont
  {P{\'e}rez-Espigares}},\ }\bibfield  {title} {\enquote {\bibinfo {title}
  {Fluctuating hydrodynamics, current fluctuations, and hyperuniformity in
  boundary-driven open quantum chains},}\ }\href@noop {} {\bibfield  {journal}
  {\bibinfo  {journal} {Physical Review E}\ }\textbf {\bibinfo {volume} {96}},\
  \bibinfo {pages} {052118} (\bibinfo {year} {2017})}\BibitemShut {NoStop}%
\bibitem [{\citenamefont {Bernard}\ and\ \citenamefont
  {Jin}(2019)}]{bernardPRL19}%
  \BibitemOpen
  \bibfield  {author} {\bibinfo {author} {\bibfnamefont {D.}~\bibnamefont
  {Bernard}}\ and\ \bibinfo {author} {\bibfnamefont {T.}~\bibnamefont {Jin}},\
  }\bibfield  {title} {\enquote {\bibinfo {title} {Open quantum symmetric
  simple exclusion process},}\ }\href@noop {} {\bibfield  {journal} {\bibinfo
  {journal} {Physical Review Letters}\ }\textbf {\bibinfo {volume} {123}},\
  \bibinfo {pages} {080601} (\bibinfo {year} {2019})}\BibitemShut {NoStop}%
\bibitem [{\citenamefont {Haga}\ \emph {et~al.}(2023)\citenamefont {Haga},
  \citenamefont {Nakagawa}, \citenamefont {Hamazaki},\ and\ \citenamefont
  {Ueda}}]{hagaPRR23}%
  \BibitemOpen
  \bibfield  {author} {\bibinfo {author} {\bibfnamefont {T.}~\bibnamefont
  {Haga}}, \bibinfo {author} {\bibfnamefont {M.}~\bibnamefont {Nakagawa}},
  \bibinfo {author} {\bibfnamefont {R.}~\bibnamefont {Hamazaki}}, \ and\
  \bibinfo {author} {\bibfnamefont {M.}~\bibnamefont {Ueda}},\ }\bibfield
  {title} {\enquote {\bibinfo {title} {Quasiparticles of decoherence processes
  in open quantum many-body systems: Incoherentons},}\ }\href@noop {}
  {\bibfield  {journal} {\bibinfo  {journal} {Physical Review Research}\
  }\textbf {\bibinfo {volume} {5}},\ \bibinfo {pages} {043225} (\bibinfo {year}
  {2023})}\BibitemShut {NoStop}%
\bibitem [{\citenamefont {Metaxas}\ \emph {et~al.}(1983)\citenamefont
  {Metaxas}, ,\ and\ \citenamefont {Meredith}}]{metaxas1983industrial}%
  \BibitemOpen
  \bibfield  {author} {\bibinfo {author} {\bibfnamefont {A.~C.}\ \bibnamefont
  {Metaxas}}, , \ and\ \bibinfo {author} {\bibfnamefont {R.~J.}\ \bibnamefont
  {Meredith}},\ }\href@noop {} {\emph {\bibinfo {title} {Industrial microwave
  heating}}},\ \bibinfo {number} {4}\ (\bibinfo  {publisher} {IET},\ \bibinfo
  {year} {1983})\BibitemShut {NoStop}%
\bibitem [{\citenamefont {Kac}(1974)}]{kac74}%
  \BibitemOpen
  \bibfield  {author} {\bibinfo {author} {\bibfnamefont {M.}~\bibnamefont
  {Kac}},\ }\bibfield  {title} {\enquote {\bibinfo {title} {{A stochastic model
  related to the telegrapher's equation}},}\ }\href@noop {} {\bibfield
  {journal} {\bibinfo  {journal} {Rocky Mountain Journal of Mathematics}\
  }\textbf {\bibinfo {volume} {4}},\ \bibinfo {pages} {497 -- 510} (\bibinfo
  {year} {1974})}\BibitemShut {NoStop}%
\bibitem [{\citenamefont {Kadanoff}\ and\ \citenamefont
  {Martin}(1963)}]{Kadanoff}%
  \BibitemOpen
  \bibfield  {author} {\bibinfo {author} {\bibfnamefont {L.~P.}\ \bibnamefont
  {Kadanoff}}\ and\ \bibinfo {author} {\bibfnamefont {P.~C.}\ \bibnamefont
  {Martin}},\ }\bibfield  {title} {\enquote {\bibinfo {title} {Hydrodynamic
  equations and correlation functions},}\ }\href@noop {} {\bibfield  {journal}
  {\bibinfo  {journal} {Annals of Physics}\ }\textbf {\bibinfo {volume} {24}},\
  \bibinfo {pages} {419--469} (\bibinfo {year} {1963})}\BibitemShut {NoStop}%
\bibitem [{\citenamefont {{\v{Z}}nidari{\v{c}}}\ and\ \citenamefont
  {Horvat}(2013)}]{vznidarivc2013transport}%
  \BibitemOpen
  \bibfield  {author} {\bibinfo {author} {\bibfnamefont {M.}~\bibnamefont
  {{\v{Z}}nidari{\v{c}}}}\ and\ \bibinfo {author} {\bibfnamefont
  {M.}~\bibnamefont {Horvat}},\ }\bibfield  {title} {\enquote {\bibinfo {title}
  {Transport in a disordered tight-binding chain with dephasing},}\ }\href@noop
  {} {\bibfield  {journal} {\bibinfo  {journal} {The European Physical Journal
  B}\ }\textbf {\bibinfo {volume} {86}},\ \bibinfo {pages} {1--11} (\bibinfo
  {year} {2013})}\BibitemShut {NoStop}%
\bibitem [{Note1()}]{Note1}%
  \BibitemOpen
  \bibinfo {note} {A Fourier transform instead of a Laplace transform can also
  be used. However, one needs to add a small negative real part to the
  eigenvalues of $A$ as the Green's function $(i\omega -A)^{-1}$ will diverge
  if $A$ has imaginary eigenvalues. To avoid this, we simply stick to the
  Laplace transform which does not have such problems.}\BibitemShut {Stop}%
\bibitem [{\citenamefont {Economou}(2006)}]{economou2006green}%
  \BibitemOpen
  \bibfield  {author} {\bibinfo {author} {\bibfnamefont {E.~N.}\ \bibnamefont
  {Economou}},\ }\href@noop {} {\emph {\bibinfo {title} {Green's functions in
  quantum physics}}},\ Vol.~\bibinfo {volume} {7}\ (\bibinfo  {publisher}
  {Springer Science \& Business Media},\ \bibinfo {year} {2006})\BibitemShut
  {NoStop}%
\bibitem [{\citenamefont {Dingfelder}\ and\ \citenamefont
  {Weideman}(2015)}]{dingfelder2015improved}%
  \BibitemOpen
  \bibfield  {author} {\bibinfo {author} {\bibfnamefont {B.}~\bibnamefont
  {Dingfelder}}\ and\ \bibinfo {author} {\bibfnamefont {J.~A.~C.}\ \bibnamefont
  {Weideman}},\ }\bibfield  {title} {\enquote {\bibinfo {title} {An improved
  {T}albot method for numerical {L}aplace transform inversion},}\ }\href@noop
  {} {\bibfield  {journal} {\bibinfo  {journal} {Numerical Algorithms}\
  }\textbf {\bibinfo {volume} {68}},\ \bibinfo {pages} {167--183} (\bibinfo
  {year} {2015})}\BibitemShut {NoStop}%
\bibitem [{\citenamefont {Cohen}(2007)}]{cohen2007numerical}%
  \BibitemOpen
  \bibfield  {author} {\bibinfo {author} {\bibfnamefont {A.~M.}\ \bibnamefont
  {Cohen}},\ }\href@noop {} {\emph {\bibinfo {title} {Numerical methods for
  {L}aplace transform inversion}}},\ Vol.~\bibinfo {volume} {5}\ (\bibinfo
  {publisher} {Springer Science \& Business Media},\ \bibinfo {year}
  {2007})\BibitemShut {NoStop}%
\bibitem [{\citenamefont {mpmath~development team}(2023)}]{mpmath}%
  \BibitemOpen
  \bibfield  {author} {\bibinfo {author} {\bibfnamefont {The}\ \bibnamefont
  {mpmath~development team}},\ }\href@noop {} {\emph {\bibinfo {title} {mpmath:
  a {P}ython library for arbitrary-precision floating-point arithmetic (version
  1.3.0)}}} (\bibinfo {year} {2023}),\ \bibinfo {note} {{\tt
  http://mpmath.org/}}\BibitemShut {NoStop}%
\bibitem [{\citenamefont {Varma}\ \emph {et~al.}(2017)\citenamefont {Varma},
  \citenamefont {de~Mulatier},\ and\ \citenamefont
  {{\v{Z}}nidari{\v{c}}}}]{varma2017fractality}%
  \BibitemOpen
  \bibfield  {author} {\bibinfo {author} {\bibfnamefont {V.~K.}\ \bibnamefont
  {Varma}}, \bibinfo {author} {\bibfnamefont {C.}~\bibnamefont {de~Mulatier}},
  \ and\ \bibinfo {author} {\bibfnamefont {M.}~\bibnamefont
  {{\v{Z}}nidari{\v{c}}}},\ }\bibfield  {title} {\enquote {\bibinfo {title}
  {Fractality in nonequilibrium steady states of quasiperiodic systems},}\
  }\href@noop {} {\bibfield  {journal} {\bibinfo  {journal} {Physical Review
  E}\ }\textbf {\bibinfo {volume} {96}},\ \bibinfo {pages} {032130} (\bibinfo
  {year} {2017})}\BibitemShut {NoStop}%
\bibitem [{\citenamefont {Ishiyama}\ \emph {et~al.}(2025)\citenamefont
  {Ishiyama}, \citenamefont {Kazuya},\ and\ \citenamefont
  {Sasamoto}}]{ishiyama2025exactdensityprofiletightbinding}%
  \BibitemOpen
  \bibfield  {author} {\bibinfo {author} {\bibfnamefont {T.}~\bibnamefont
  {Ishiyama}}, \bibinfo {author} {\bibfnamefont {F.}~\bibnamefont {Kazuya}}, \
  and\ \bibinfo {author} {\bibfnamefont {T.}~\bibnamefont {Sasamoto}},\
  }\bibfield  {title} {\enquote {\bibinfo {title} {Exact density profile in a
  tight-binding chain with dephasing noise},}\ }\href@noop {} {\  (\bibinfo
  {year} {2025})},\ \Eprint {http://arxiv.org/abs/2501.07095} {arXiv:2501.07095
  [cond-mat.stat-mech]} \BibitemShut {NoStop}%
\end{thebibliography}%
\end{document}